\documentclass[10pt]{article}
\pdfoutput=1
\usepackage{graphicx}
\usepackage{xspace}
\usepackage{mciteplus}
\usepackage{hyperref}
\usepackage{subcaption}
\usepackage{amsmath}
\usepackage{rotating}
\usepackage{float}
\usepackage{lineno}
\def\Title#1{\begin{center} {\Large #1 } \end{center}}
\def\Author#1{\begin{center}{ \sc #1} \end{center}}
\def\Address#1{\begin{center}{ \it #1} \end{center}}

\newcommand\pubblock{\rightline{\begin{tabular}{l} Proceedings of the Fifth Annual LHCP\\ \pubnumber\\
         \pubdate  \end{tabular}}}

\newenvironment{Abstract}{\begin{quotation} \begin{center} 
             \large ABSTRACT \end{center}\bigskip 
      \begin{center}\begin{large}}{\end{large}\end{center} \end{quotation}}

\newenvironment{Presented}{\begin{quotation} \begin{center} 
             PRESENTED AT\end{center}\bigskip 
      \begin{center}\begin{large}}{\end{large}\end{center} \end{quotation}}

\def\beq{\begin{equation}}
\def\eeq#1{\label{#1}\end{equation}}
\def\eeqn{\end{equation}}

\def\beqa{\begin{eqnarray}}
\def\eeqa#1{\label{#1}\end{eqnarray}}
\def\eeqan{\end{eqnarray}}

\let\bar=\overbar

\def\Dslash{\not{\hbox{\kern-4pt $D$}}}
\def\dslash{\not{\hbox{\kern-2pt $\del$}}}

\def\BR{\mbox{\rm BR}}

\def\msb{{\bar{\ssstyle M \kern -1pt S}}}

\newcommand{\Bp}{\ensuremath{B^+}\xspace}
\newcommand{\Bz}{\ensuremath{B^0}\xspace}
\newcommand{\Bd}{\ensuremath{B^0}\xspace}

\newcommand{\Bs}{\ensuremath{B^0_s}\xspace}
\newcommand{\Bc}{\ensuremath{B^0_c}\xspace}
\def\kstar{\ensuremath{K^{*}}\xspace}

\newcommand{\mumu}{\ensuremath{\mu^+\mu^-}\xspace}
\newcommand{\Bsmumu}{\ensuremath{B^0_s \to \mu^+\mu^-}\xspace}
\newcommand{\Bdmumu}{\ensuremath{B^0 \to \mu^+\mu^-}\xspace}
\newcommand{\Bmumu}{\ensuremath{B^0{}\!_{(s)} \to \mu^+\mu^-}\xspace}

\newcommand{\BpKpJpsi}{\ensuremath{B^+ \to J/\psi \,K^+}\xspace}

\newcommand{\Bhh}{\ensuremath{B^0{}\!_{(s)} \to hh^\prime}\xspace}

\newcommand{\theInvRho}{\ensuremath{\frac{\varepsilon_{\mu^+\mu^-}}{\varepsilon_{J/\psi \, K^+}}}\xspace}

\def\bdkstmumu{\ensuremath{\Bd \to K^{*}\mumu}\xspace}
\def\afb{\ensuremath{A_{\mathrm{FB}}}\xspace}
\def\qsq{\ensuremath{q^2}\xspace}
\def\gev{\ensuremath{\mathrm{GeV}}\xspace}
\def\qsqunits{\ensuremath{\gev^2}\xspace}
\def\invfb{\ensuremath{\mathrm{fb}^{-1}}\xspace}
\def\jpsi{\ensuremath{J/\psi}\xspace}
\def\psitwos{\ensuremath{\psi(2S)}\xspace}
\def\LambdaB{\ensuremath{\Lambda_b}\xspace}
\def\tk{\ensuremath{\theta_{{K}}}\xspace}
\def\tl{\ensuremath{\theta_{{L}}}\xspace}
\def\ctk{\ensuremath{\cos\theta_{{K}}}\xspace}
\def\ctl{\ensuremath{\cos\theta_{{L}}}\xspace}
\def\ctksq{\ensuremath{\cos^2\theta_{{K}}}\xspace}

\def\stl{\ensuremath{\sin\theta_{{L}}}\xspace}
\def\stksq{\ensuremath{\sin^2\theta_{{K}}}\xspace}
\def\stlsq{\ensuremath{\sin^2\theta_{{L}}}\xspace}

\def\fl{\ensuremath{F_L}\xspace}

\def\afb{\ensuremath{A_{\mathrm{FB}}}\xspace}

\def\Sthree{\ensuremath{S_3}\xspace}
\def\Sfour{\ensuremath{S_4}\xspace}
\def\Sfive{\ensuremath{S_5}\xspace}
\def\Ssix{\ensuremath{S_6}\xspace}

\def\Sseven{\ensuremath{S_7}\xspace}
\def\Seight{\ensuremath{S_8}\xspace}
\def\Snine{\ensuremath{S_9}\xspace}

\def\Pone{\ensuremath{P_1}\xspace}

\def\Pfour{\ensuremath{P_4^\prime}\xspace}
\def\Pfive{\ensuremath{P_5^\prime}\xspace}
\def\Psix{\ensuremath{P_6^{\prime}}\xspace}

\def\Peight{\ensuremath{P_8^\prime}\xspace}

\newcommand{\qsqrbin}[2]{\ensuremath{[#1, #2]}\xspace}

\def\qsqrbinA{\qsqrbin{0.04}{2.0}}
\def\qsqrbinB{\qsqrbin{2.0}{4.0}}
\def\qsqrbinC{\qsqrbin{4.0}{6.0}}
\def\qsqrbinD{\qsqrbin{0.04}{4.0}}
\def\qsqrbinE{\qsqrbin{1.1}{6.0}}
\def\qsqrbinF{\qsqrbin{0.04}{6.0}}
\def\qsqrbinning{\qsqrbinA, \qsqrbinB, \qsqrbinC, \qsqrbinD, \qsqrbinE, \qsqrbinF}

\usepackage{color}

\newcommand\pubnumber{ ATL-PHYS-PROC-2017-233 }

\newcommand\pubdate{\today}

\def\affiliation{
On behalf of the ATLAS Collaboration, \\
School of Physics and Astronomy \\
Queen Mary University of London, London, UK}

\begin{document}

\large
\begin{titlepage}
\pubblock

\vfill
\Title{Study of $b \to s\ell\ell$ decays at ATLAS}
\vfill

\Author{ Marcella Bona }
\Address{\affiliation}
\vfill
\begin{Abstract}

  The study of flavour-changing neutral currents (FCNC) gives access to important tests
  of the Standard Model (SM) and allows to search for hints of beyond the SM phenomena.
  We present here the study of the very rare decays $B^0\to \mu^+\mu^-$ and $B_s^0\to \mu^+\mu^-$
  using data corresponding to an integrated luminosity of 25 fb$^{-1}$ of $7$~TeV and $8$~TeV
  proton--proton collisions collected with the ATLAS detector during the LHC Run 1.
  For $B^0$, an upper limit on the branching fraction is set at BR$(B^0\to\mu^+\mu^-)<4.2\times 10^{-10}$
  at 95\% confidence level. For $B^0_s$, the branching fraction
  BR$(B^0_s\to\mu^+\mu^-)=(0.9^{+1.1}_{-0.8})\times10^{-9}$ is measured.
  The results are consistent with the SM expectation with a
  p-value of 4.8\%, corresponding to 2.0 standard deviations.
  Another study sensitive to possible new physics contributions in $b \to s\ell\ell$ decays
  is the angular analysis of the decay $B^0\to K^*\mu^+\mu^-$. Here we present the results
  obtained using proton--proton collisions at $\sqrt{s}=8$~TeV from LHC data collected
  with the ATLAS detector. The study is based on 20.3 fb$^{-1}$ of integrated luminosity collected
  during 2012. Measurements of the $K^*$ longitudinal polarisation fraction and a set of
  angular parameters obtained for this decay are presented. The results are compared
  to a variety of theoretical predictions and found to be compatible with them.

\end{Abstract}
\vfill

\begin{Presented}
The Fifth Annual Conference\\
on Large Hadron Collider Physics \\
Shanghai Jiao Tong University, Shanghai, China\\ 
May 15-20, 2017
\end{Presented}
\vfill
\end{titlepage}
\def\thefootnote{\fnsymbol{footnote}}
\setcounter{footnote}{0}
%

\normalsize 


\section{Introduction}

Flavor-changing neutral currents (FCNC) have played a significant role in
the construction of the Standard Model of particle physics (SM).
These processes are forbidden at tree level in the SM, hence they need to proceed
at next-to-leading order, via loops, resulting in being rare or even very rare.
An important set of FCNC processes involve the transition of
a $b$ quark to an $s\mumu$ final state mediated by electroweak box and
loop diagrams. If heavy new particles exist, they may contribute
to FCNC decay amplitudes, affecting the measurement of SM observables.
The branching fractions of the $\Bmumu$ decays are of particular interest
because the additional helicity suppression makes them very rare and
because they are accurately predicted in the SM.
Also the decay $\Bd\to \kstar(892)\mu^+\mu^-$, where $\kstar(892)\to K^+\pi^-$,
and in particular the angular distribution of its four-body final state
gives access to numerous observables sensitive to new physics.
Angular observables such as the forward-backward asymmetry (\afb) should
be measured as a function of the invariant mass squared of the di-lepton
system (\qsq), as they can be sensitive to different types of new
physics introduced as FCNCs at loop level.

\section{Study of the rare decays of $B^0_s$ and $B^0$ into muon pairs}

We present here the result of a search for \Bsmumu\ and \Bdmumu\ decays~\cite{BPHY-2012-01}
performed using $pp$ collision data corresponding to an integrated
luminosity of 25~fb$^{-1}$, collected at $7$ and $8$~TeV
in the full LHC Run~1 data-taking period using the ATLAS detector~\cite{AtlasDet}.
The SM predictions for $\Bmumu$ decays are
$\BR(\Bsmumu) = (3.65 \pm 0.23) \times 10^{-9}$ and
$\BR(\Bdmumu) = (1.06 \pm 0.09) \times 10^{-10}$~\cite{Bobeth:2013uxa}.
The LHCb collaboration has recently reported a first single-experiment observation
of \Bsmumu obtaining $\BR(\Bsmumu) = \left(2.8^{+0.7}_{-0.6}\right) \times 10^{-9}$ and
$\BR(\Bdmumu) = \left(3.9^{+1.6}_{-1.4}\right) \times 10^{-10}$~\cite{Aaij:2017vad},
while the first observation has been obtained by the combined analyses of the LHCb and
CMS collaborations~\cite{CMS:2014xfa}.

The \Bsmumu\ and \Bdmumu\ branching fractions are measured relative
to the normalisation decay $B^+\to J/\psi (\to \mu^+\mu^-)K^+$ that is
abundant and has a known branching fraction.
The \Bdmumu\ (\Bsmumu) branching fraction can be extracted as:
\begin{equation}
  \BR (B^0_{(s)}\! \to \!\mu^+ \mu^-)
  = N_{d(s)} \times \left[\BR (\BpKpJpsi) \times \BR (J/\psi \to \mumu)\right]
  \times \frac{f_{u}}{f_{d(s)}}
  \times \frac{1}{\mathcal{D}_{\mathrm{norm}}}\,,  \label{eq:BRFormula} \\
\end{equation}
\vspace*{-0.3cm}
\begin{equation}
\mbox{with}\;\;\;\;\;\;\;\;\;
  \mathcal{D}_{\mathrm{norm}} =  \sum_k N^k_{J/\psi K^+} \alpha_k\left(\theInvRho\right)_k \, ,
  \label{eq:DnormFormula}
\end{equation}
where $N_d$ ($N_s$) is the \Bdmumu\ (\Bsmumu) signal yield,
$N_{J/\psi K^+}$ is the \BpKpJpsi\ normalisation yield,
$\varepsilon_{\mu^+\mu^-}$ and $\varepsilon_{J/\psi K^+}$
are the values of acceptance times efficiency,
and $f_u/f_d$  ($f_u/f_s$) is the ratio of the hadronisation probabilities of a
$b$-quark into \Bp\ and \Bz\ (\Bs).
The denominator $\mathcal{D}_{\mathrm{norm}}$ consists of a sum
whose index $k$ runs over the data-taking periods and the trigger selections.
The $\alpha_k$ parameter takes into account the different trigger
prescale factors and integrated luminosities in the signal and normalisation channels.
The ratio of the efficiencies corrects for reconstruction
differences in each data sample $k$.
Signal and reference channel events are selected with similar di-muon triggers.

The background to the \Bmumu\ signal originates from three main sources:
\vspace*{-0.2cm}
\begin{description}
  \setlength{\itemsep}{0pt}%
  \setlength{\parskip}{0pt}%
\item[] {\em continuum} background, the dominant combinatorial component,
  made from muons coming from uncorrelated hadron decays and characterised
  by a smooth di-muon invariant mass distribution.
  It is studied in the signal mass sidebands, and in an inclusive MC sample
  of semileptonic decays of $b$ and $c$ hadrons.
\item[] {\em partially reconstructed}  $B\to\mu\mu X$ decays, characterised
  by non-reconstructed final-state particles ($X$) and thus accumulating
  in the low di-muon invariant mass sideband;
\item[] { \em peaking} background, due to $B^0{}\!_{(s)}\to h\,h^\prime$ decays,
  with both hadrons misidentified as muons.
\end{description}
\vspace*{-0.2cm}

\begin{figure}[!tb]
  \begin{center}
    \hspace*{-0.6cm}
    \begin{subfigure}[t]{0.36\textwidth}
      \includegraphics[width=0.99\textwidth]{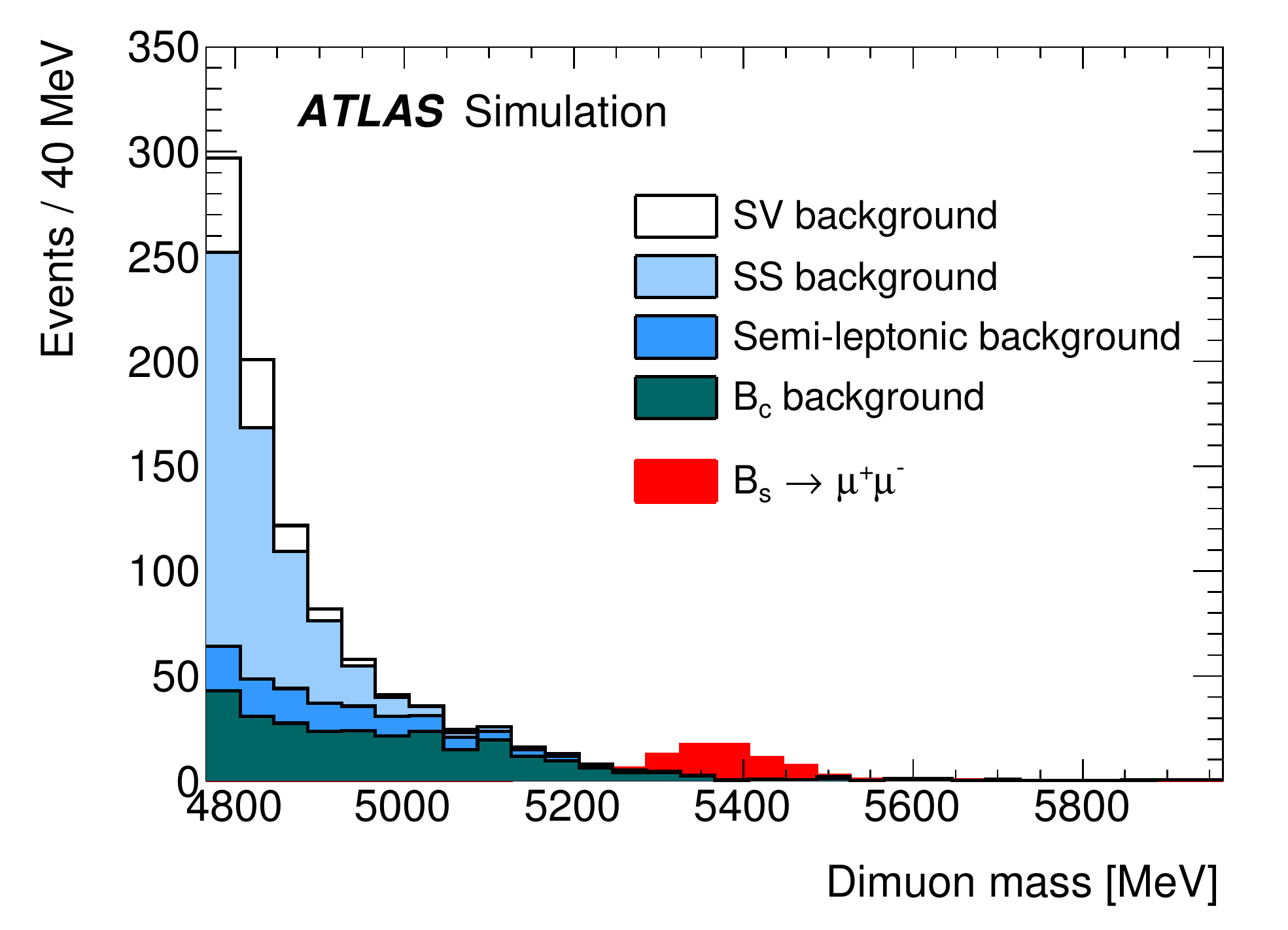}
    \end{subfigure}
    \hspace*{-0.6cm}
    \begin{subfigure}[t]{0.36\textwidth}
      \includegraphics[width=0.99\textwidth]{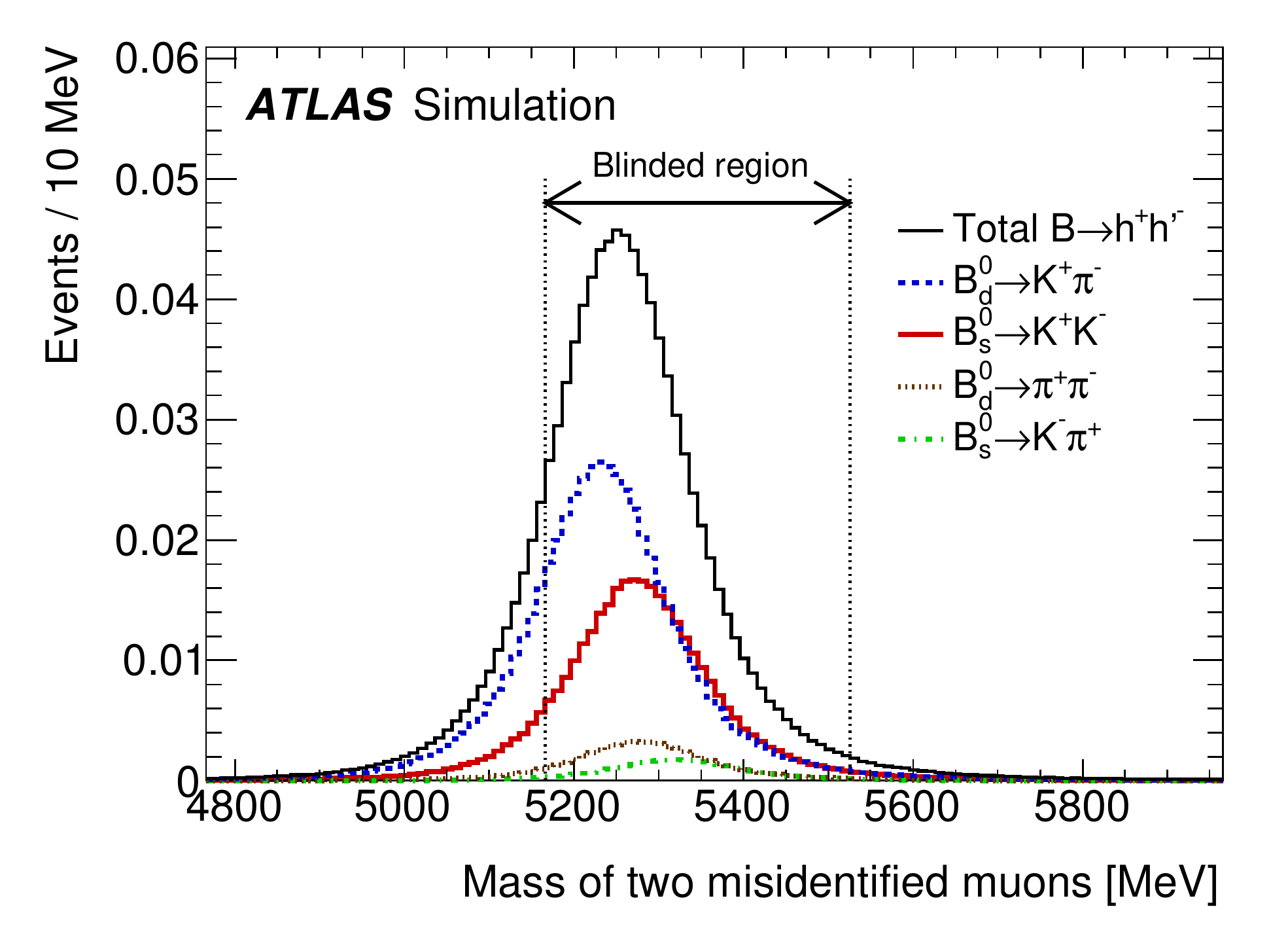}
    \end{subfigure}
    \hspace*{-0.6cm}
    \begin{subfigure}[t]{0.36\textwidth}
      \includegraphics[width=0.99\textwidth]{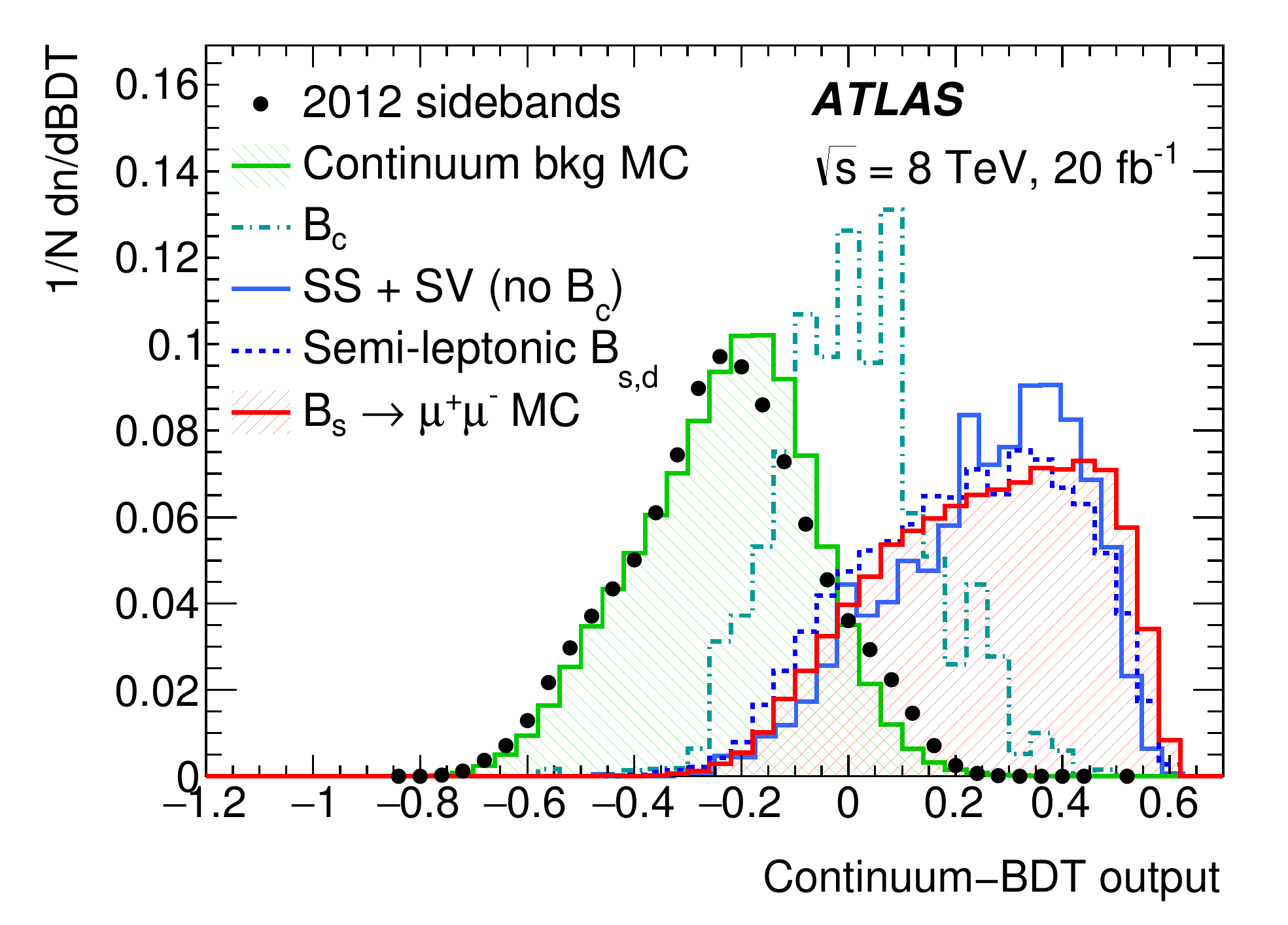}
    \end{subfigure}
    \hspace*{-0.8cm}
    \vspace*{-0.2cm}
    \caption{{\it{Left:}} di-muon invariant mass distribution from simulation
      for partially reconstructed backgrounds.
      Components are shown as stacked histograms, normalised according
      to world-averaged branching fractions. The SM expectation for the
      \Bsmumu\ signal is also shown for comparison (non-stacked).
      {\it{Centre:}} invariant mass distribution for the peaking background \Bhh.
      {\it{Right:}} continuum-BDT distribution for signal and background
      events: signal $B^0{}\!_{(s)}$, partially reconstructed $B$ events (SS+SV),
      \Bc\ decays and continuum. The solid histograms are obtained from simulation,
      while the points represent data collected in the sidebands~\cite{BPHY-2012-01}.}
    \label{fig:bkg}
  \end{center}
  \vspace*{-0.5cm}
\end{figure}

\noindent The partially reconstructed decays consist of several topologies:
(a) {\em same-side} (SS) combinatorial background from decay cascades
($b \to c \mu\nu \to  s(d) \mu\mu\nu\nu$);
(b) {\em same-vertex} (SV) background from $B$ decays containing a muon pair
(e.g. $B^0 \to K^{*0}\mu\mu$, $B \to J/\psi X \to \mu\mu\mu X^\prime$);
(c) $B_c$ decays (e.g. $B_c \to J/\psi \mu \nu \to \mu\mu\mu\nu$);
(d) semileptonic $b$-hadron decays where a final-state hadron is
misidentified as a muon. The latter are mainly three-body charmless decays
$B^0 \to \pi\mu\nu$, $B^0_s \to K\mu\nu$ and $\Lambda_b \to p\mu\nu$ and
their contribution is reduced by the muon identification requirements.
The MC invariant mass distributions of all these topologies are shown
in the left plot in Figure~\ref{fig:bkg}.
The peaking background is due to $B^0{}\!_{(s)}$ decays containing
two hadrons misidentified as muons, which populate the signal region
as shown in the central plot in Figure~\ref{fig:bkg}.

Two multivariate discriminants, implemented as BDTs, have been employed:
the ``fake-BDT'' aims at minimising the amount of hadrons erroneously
identified as muons and the ``continuum-BDT'' aims at discriminating
against the continuum background.
For the fake-muon background, the vast majority of events with hadron
misidentification are due to decays in flight of kaons and pions.
The fake-BDT selection is tuned for a 95\% efficiency for signal muons,
and the resulting misidentification probability is equal to 0.09\% for
kaons and 0.04\% for pions.
The proton misidentification rate is negligible ($<$ 0.01\%).
After the fake-BDT selection, the expected number of peaking-background events
is equal to 0.7, with 10\% uncertainty.

The continuum-BDT is based on $15$ variables exploiting the reconstruction
of the $B$ decay vertex, the separation between production and decay vertices, and
the characteristics of the signal muons and of the rest of the event.
Right plot in Figure~\ref{fig:bkg} shows the distribution of the continuum-BDT variable
for signal and the various backgrounds.
The final selection requires a continuum-BDT value larger than $0.24$, corresponding
to a signal relative efficiency of $54\%$, and to a reduction of the continuum background
by a factor of $\sim 10^{-3}$.

\begin{figure}[!tb]
  \begin{center}
    \hspace*{-0.6cm}
    \begin{subfigure}[t]{0.36\textwidth}
      \includegraphics[width=0.99\textwidth]{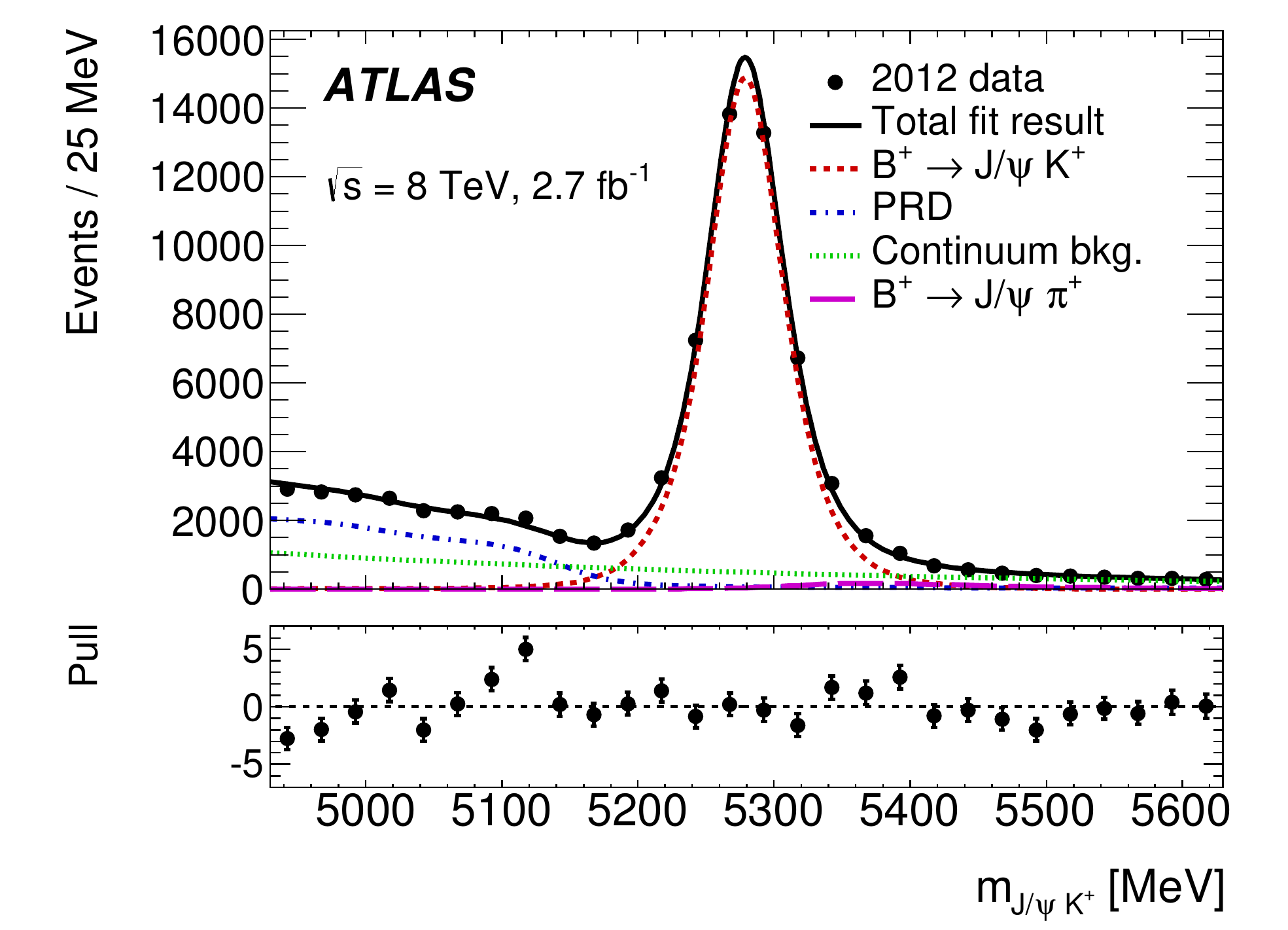}
    \end{subfigure}
    \hspace*{-0.6cm}
    \begin{subfigure}[t]{0.36\textwidth}
      \includegraphics[width=0.99\textwidth]{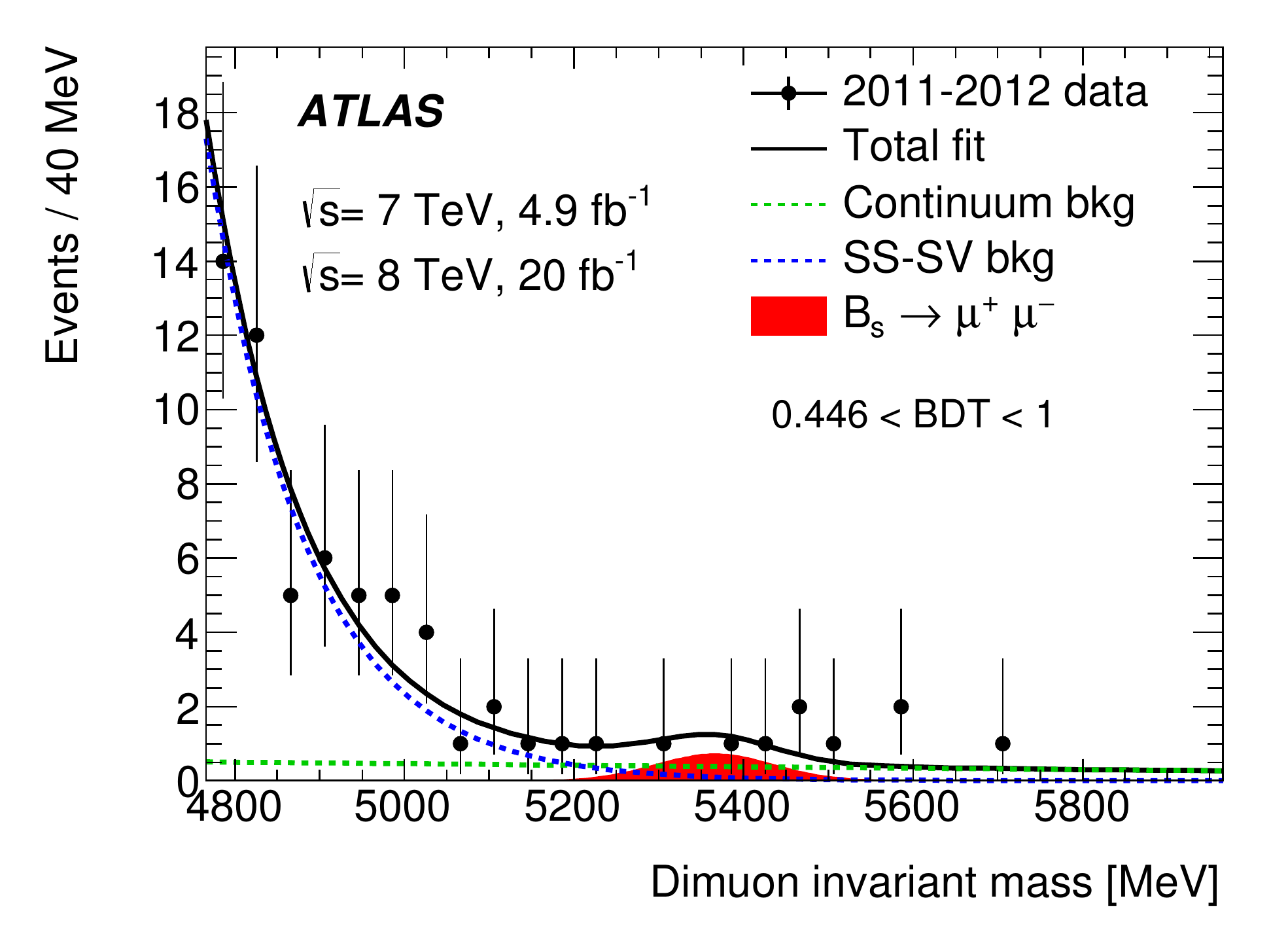}
    \end{subfigure}
    \hspace*{-0.6cm}
    \begin{subfigure}[t]{0.36\textwidth}
      \includegraphics[width=0.99\textwidth]{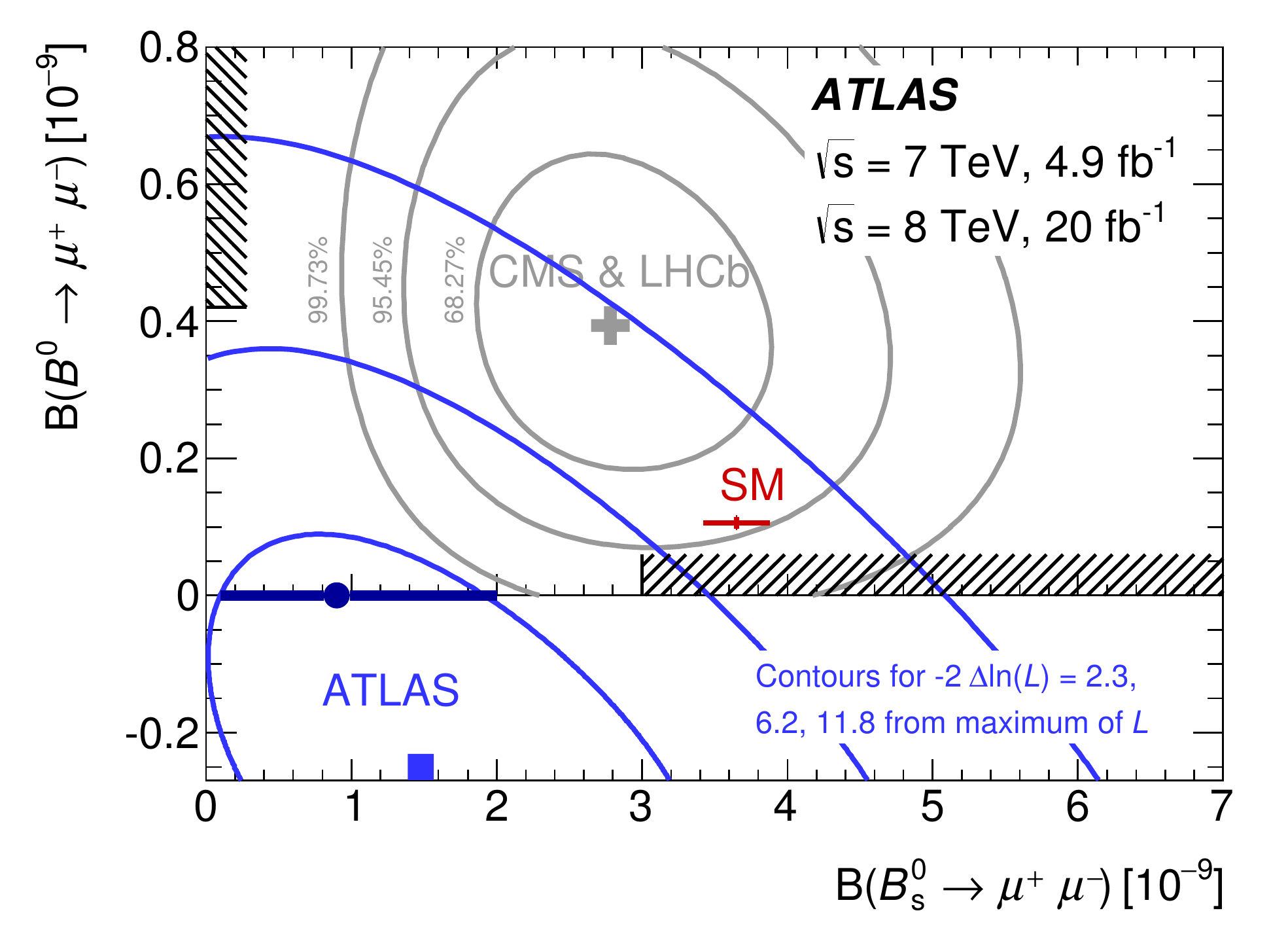}
    \end{subfigure}
    \hspace*{-0.8cm}
    \vspace*{-0.2cm}
    \caption{{\it{Left:}} $J/\psi K^+$ invariant mass distribution for $B^+$
      candidates in the main trigger category in 2012 data. The result of the
      fit is overlaid. {\it{Centre:}} di-muon invariant mass distribution in
      Run~1 data, in the third continuum-BDT interval. The result of the fit
      for non-negative signal contributions is overlaid. {\it{Right:}} contours in the
      $\BR (\Bsmumu), \BR (\Bdmumu)$ plane for intervals of $-2\, \Delta \ln(L)$ equal
      to $2.3$, $6.2$ and $11.8$ relative to the absolute maximum of the likelihood,
      without the constraint of non-negative branching fractions.
      The hatched areas on the axes correspond to the excluded values from the
      upper limits at 95\% CL. Also shown are the contours for the combined CMS-LHCb result,
      the SM prediction, and the maximum of the likelihood with non-negative
      branching fractions~\cite{BPHY-2012-01}.
    }
    \label{fig:sig}
  \end{center}
  \vspace*{-0.5cm}
\end{figure}

The $B^+$ yields are extracted with unbinned extended maximum-likelihood fits to the
$J/\psi K^+$ invariant mass distributions. An example is shown in the
left plot in Figure~\ref{fig:sig}.
All the yields are extracted from fits to data, while the shape parameters
are determined from a simultaneous fit to data and MC samples.
Free parameters are introduced for the mass scale and mass resolution
to accommodate data--MC differences.

Both \BpKpJpsi\ and \Bmumu\ channels are measured in the fiducial volume of the $B$
meson defined as $p_{\text{T}}^B>8.0$~GeV and $\left| \eta_B\right|<2.5$.
The total efficiencies within the fiducial volume include acceptance and trigger,
reconstruction and selection efficiencies.
All efficiency terms are computed on data-corrected MC samples separately for the
three trigger selections used in 2012 and for the 2011 sample.
The ratios of efficiencies enters in Eq.~(\ref{eq:DnormFormula}):
their statistical uncertainties come from the finite size of the MC samples.
The systematic uncertainties come from data corrections, trigger efficiencies,
the data--MC differences, and the differences between the \Bsmumu\ and the \BpKpJpsi\ channels.
The total uncertainty on $\mathcal{D}_{\mathrm{norm}}$ is $\pm 5.9$\%.
A correction to the efficiency for \Bsmumu\ is needed because of the width difference
$\Delta \Gamma_s$ between the \Bs\ eigenstates: the variation in the \Bsmumu\ mean lifetime
changes the \Bs\ efficiency by $+4$\%.

The total yields of \Bsmumu\ and \Bdmumu\ events are obtained from an unbinned extended
maximum-likelihood fit on the di-muon invariant mass distribution performed simultaneously across
three continuum-BDT intervals. Each interval corresponds to an equal efficiency of $18\%$
for signal events, and it is ordered according to increasing signal-to-background ratio.
The central plot of Figure~\ref{fig:sig} shows the di-muon invariant mass distributions
in the Run~1 data in the third interval.
The values determined by the fit are $N_s = 16 \pm 12$ and $N_d = - 11 \pm 9$.
The systematic uncertainties related to the continuum-BDT intervals and to the 
the fit models are included in the likelihood with Gaussian multiplicative
factors with width equal to the systematic uncertainty value.
The primary result of this analysis is obtained by applying the natural boundary of
non-negative yields, for which the fit returns the values $N_s = 11$ and $N_d = 0$.

The branching fractions for the decays \Bsmumu\ and \Bdmumu\ are extracted
using a profile-likelihood fit. The likelihood is obtained from the one
used for the yields replacing the fit parameters with the corresponding branching
fractions divided by normalisation terms in Eq.~(\ref{eq:BRFormula}), and including
Gaussian multiplicative factors for the normalisation uncertainties.
A Neyman construction~\cite{neyman} is used to determine the $68.3\%$ confidence interval
for \BR(\Bsmumu) with pseudo-MC experiments, obtaining:
\begin{equation}
  \BR (\Bsmumu) =  \left( 0.9^{+1.1}_{-0.8} \right) \times 10^{-9}\, ,
\end{equation}
where the uncertainties are both statistical and systematic.
The statistical uncertainty is dominant, with the systematic uncertainty equal to
$\pm \,0.3\times 10^{-9}$.
The observed significance of the \Bsmumu\ signal from pseudo-MC experiments
is equal to 1.4 standard deviations ($\sigma$).
The expected significance for the SM prediction~\cite{Bobeth:2013uxa}
is $3.1\sigma$.
Pseudo-MC experiments are used to evaluate the compatibility with the SM prediction:
for the simultaneous fit to \BR(\Bsmumu) and \BR(\Bdmumu),
the result is $p=0.048\pm0.002$, corresponding to $2.0 \sigma$.
The right plot in Figure~\ref{fig:sig} shows the contours in the plane of \BR(\Bsmumu)
and \BR(\Bdmumu) for values of $-2\, \Delta \ln(L)$ equal to $2.3$, $6.2$ and $11.8$,
relative to the maximum of the likelihood, allowing negative values of the branching
fractions. The maximum within the physical boundary is shown with error bars
corresponding to the $68.3$\% interval for \BR(\Bsmumu).
With the CL$_\mathrm{s}$ method~\cite{Read:2002hq}, upper limits are placed on both
the \Bsmumu\ and \Bdmumu\ branching fractions at the $95\%$ confidence level:
$\BR(\Bsmumu) < 3.0\times 10^{-9}$ and $\BR(\Bdmumu) < 4.2\times 10^{-10}$.
The expected significance for \BR(\Bdmumu) SM prediction is equal to
$0.2 \sigma$.

\section{Angular analysis of \bdkstmumu decays}

We present here the results of the angular analysis of the decay \bdkstmumu
with the ATLAS detector, using 20.3~\invfb of $pp$ collision data at a
centre of mass energy $\sqrt{s}=8$~TeV delivered by the LHC during 2012~\cite{ATLAS-CONF-2017-023}.
In order to compare with other experiments and phenomenology studies,
results are presented in six different bins of \qsq\ in the range $0.04$
to $6.0$ \qsqunits, where three of these bins overlap.

Three angular variables are used to describe the decay:
the angle between the $K^+$ and the direction opposite to the \Bz in the $K^*$ centre
of mass frame (\tk); the angle between the $\mu^+$ and the direction opposite to the
\Bz in the di-muon centre of mass frame (\tl); and the angle between the two decay
planes formed by the $K\pi$ and the di-muon systems in the \Bz rest frame ($\phi$).
Figure~\ref{fig:angles} illustrates the angles used.
The angular differential decay rate for \bdkstmumu is a function of \qsq, \ctk, \ctl\ and
$\phi$ is expressed as a function of the angular parameters
via coefficients that may be represented by the helicity or transversity
amplitudes and is written as:\\
\hspace*{-1.3cm}
\begin{minipage}{0.65\textwidth}
  \begin{eqnarray}
    \footnotesize
    && \!\frac{1}{\mathrm{d}\Gamma / \mathrm{d}\qsq}\frac{\mathrm{d}^4\Gamma}{\mathrm{d}\ctl
      \mathrm{d}\ctk \mathrm{d}\phi \mathrm{d}\qsq} = \nonumber \\
    && \;\;\;\;\;\;= \frac{9}{32\pi}\Bigg[\frac{3(1-\fl)}{4}\stksq \nonumber \\
      && \;\;\;\;\;\;+ \fl \ctksq + \frac{1-\fl}{4} \stksq \cos 2\theta_L \nonumber \\
      && \;\;\;\;\;\;- \fl \ctksq\cos 2\theta_L + \Sthree \stksq\stlsq\cos 2 \phi \nonumber\\
      && \;\;\;\;\;\;+ \Sfour \sin 2\theta_K \sin 2\theta_L \cos\phi \nonumber + \Sfive\sin 2\theta_K\stl\cos\phi\nonumber \\
      && \;\;\;\;\;\;+ \Ssix \stksq \ctl + \Sseven\sin 2\theta_K\stl\sin\phi \nonumber \\
      && \;\;\;\;\;\;+ \Seight\sin 2\theta_K\sin 2\theta_L \sin\phi + \Snine \stksq\stlsq\sin 2\phi \Bigg]. \label{eq:fullangcorrelation}
  \end{eqnarray}
\end{minipage}
\hspace*{1.5cm}
\begin{minipage}{0.32\textwidth}
  \begin{figure}[H]
    \includegraphics[width=1.\textwidth]{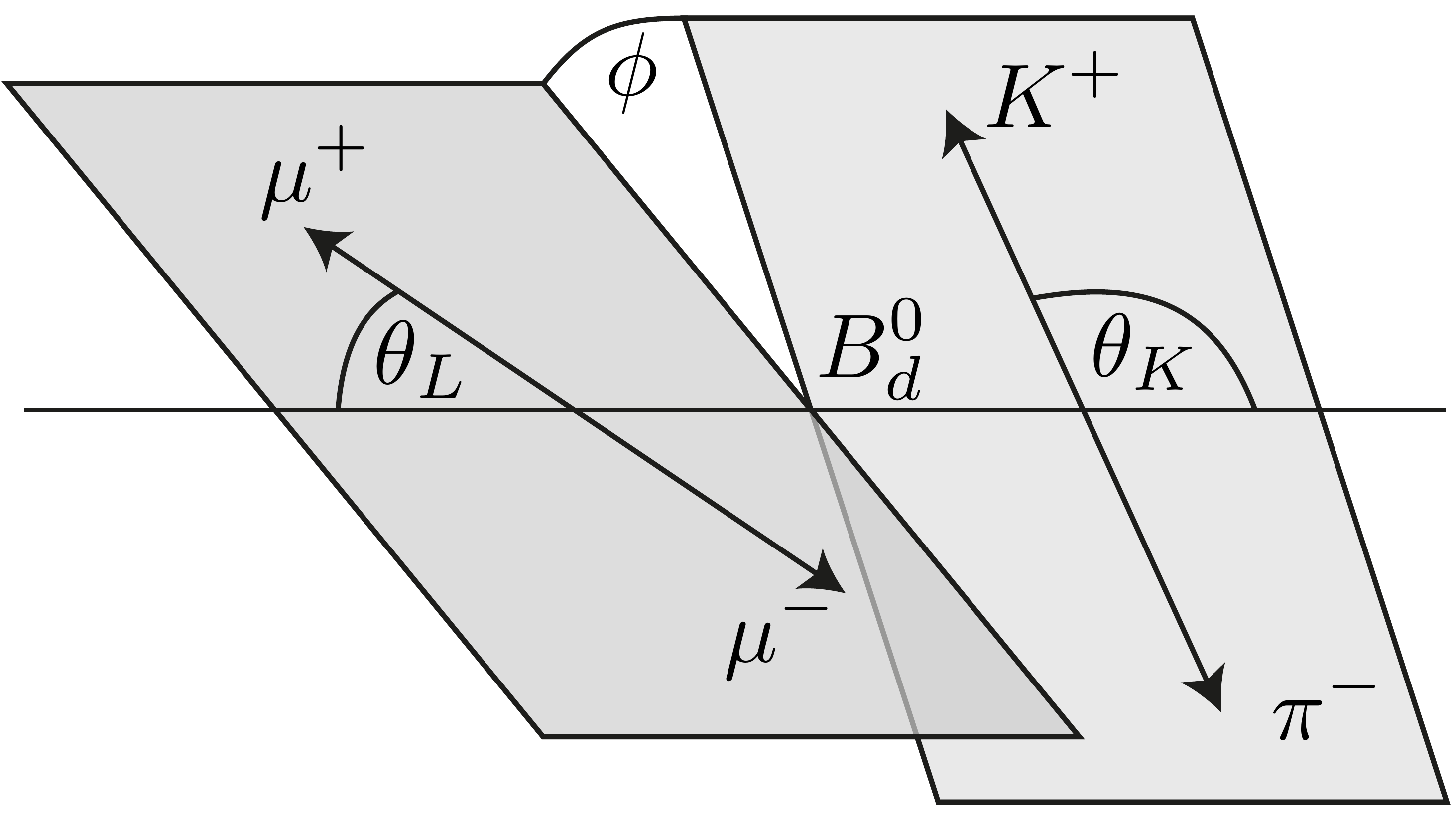}
    \caption{Illustration of the \bdkstmumu decay showing the angles
      $\theta_K$, $\theta_L$ and $\phi$ defined in the text.
      Angles are computed in the rest frame of the $K^*$, di-muon system and
      \Bz meson, respectively~\cite{ATLAS-CONF-2017-023}.}\label{fig:angles}
  \end{figure}
\end{minipage}\vspace*{0.5cm}\\
\noindent Here $\fl$ is the fraction of longitudinally polarised $\kstar$s and the $S_i$ are
angular coefficients.  These angular parameters are functions of the real and
imaginary parts of the transversity amplitudes of $\Bd$ decays to $\kstar\mu^+\mu^-$.
The forward-backward asymmetry is given by $\afb = 3\Ssix/4$.
The $S_i$ parameters depend on hadronic form factors which have significant uncertainties
at leading order. It is possible to reduce the theoretical uncertainty on the parameters
extracted from data by transforming the $S_i$ using ratios constructed to cancel form
factor uncertainties at leading order.
These ratios are given by Refs~\cite{DescotesGenon:2012zf,*Descotes-Genon:2013vna} as
\begin{equation}
  P_1 = \frac{2\Sthree}{1-\fl}  \;\;\;\;\;\;\;\;
  P_2 = \frac{2}{3}\frac{\afb}{1-\fl}  \;\;\;\;\;\;\;\;
  P_3 = -\frac{\Snine}{1-\fl}  \;\;\;\;\;\;\;\;
  P_{j=4,5,6,8}^{\prime} = \frac{S_{i=4,5,7,8}}{\sqrt{\fl(1-\fl)}}.
\end{equation}
All of the parameters introduced, \fl, $S_i$ and $P^{(\prime)}_j$, vary with \qsq\ and
the data are analysed in \qsq\ bins to obtain an average value for a given parameter
in that bin.  Measurements of these quantities can be used as inputs to global fits
used to determine the values of Wilson coefficients and search for new physics.

As the ATLAS detector does not have a dedicated charged-particle identification system,
$\kstar$ candidates are reconstructed to satisfy both possible $K\pi$ mass hypotheses.
The invariant $K\pi$ mass is required to lie in the range $[846, 946]$~MeV.
The charge of the kaon candidate is used to assign the flavor of the $B$ candidate.
This procedure results in an incorrect flavor tag (mistag $\omega$) of $\omega = 0.1088\pm 0.0005$
($\overline \omega = 0.1086\pm 0.0005$).

The region $\qsq\in[0.98, 1.1]$ \qsqunits is vetoed to remove any contamination from
the $\phi(1020)$ resonance. The remaining data with $\qsq \in [0.04, 6.0]$ \qsqunits are analysed.
Two $K^*c\overline c$ control sample regions are defined for $B$ decays to $K^*\jpsi$ and
$K^*\psitwos$, respectively as $\qsq\in[8, 11]$ and $[12, 15]$ \qsqunits.
The control samples are used to extract nuisance parameters of the signal probability density
function (p.d.f.) from data.
For \qsq $<6$ \qsqunits the selected data sample consists of $787$ events and is composed of
signal $B^0_d \to K^*\mumu$ decay events as well as background that is dominated by a
combinatorial component that does not peak in $m_{K\pi\mu\mu}$ and does not exhibit
a resonant structure in \qsq.
Above $6$ \qsqunits several backgrounds pass the selection imposed, including events coming
from the low mass tail of $B\to K^*\jpsi$.
The data are analysed in the \qsq bins \qsqrbinning \qsqunits,
where the bin width is chosen to provide a sample of signal events sufficient to perform an
angular analysis and to be larger than the mass resolution.

Extended unbinned maximum likelihood fits of the angular distributions of the decay
are performed on the data for each \qsq bin. The variables used in the fit
are $m_{K\pi\mu\mu}$, the cosines of the helicity angles ($\ctk$ and $\ctl$), and $\phi$.
The full angular distribution cannot be reliably fit given the current statistics:
trigonometric transformations are used to simplify the distribution through `folding schemes'.
Each transformation simplifies Eq.~(\ref{eq:fullangcorrelation}) such that only three
parameters are extracted from each fit: \fl, \Sthree and one of the other $S_i$ parameters.
The values and uncertainties of \fl and \Sthree obtained from the four fits are consistent
with each other and the results reported are those found to have the smallest systematic
uncertainty. This procedure implies that \Ssix (\afb) and \Snine cannot be extracted.
The angular parameters of interest for these schemes are $(\fl, \Sthree, S_i)$
where $i=4, 5, 6, 8$. These translate into $(\fl, P_1, P^\prime_j)$, where $j=4, 5, 7, 8$.

\begin{table*} [!tb]
  \small
  \begin{center}
    \begin{tabular}{l|cc}\hline\hline
      \qsq [$\qsqunits$] & $n_\mathrm{signal}$      & $n_\mathrm{background}$  \\ \hline
      \qsqrbinA          & $128 \pm 22$ & $122 \pm 22$  \\
      \qsqrbinB          & $106 \pm 23$ & $113\pm 23$\\
      \qsqrbinC          & $114\pm 24$ & $204\pm 26$\\
      \hline\hline
    \end{tabular}
    \hspace{0.5cm}
    \begin{tabular}{l|cc}\hline\hline
      \qsq [$\qsqunits$] & $n_\mathrm{signal}$      & $n_\mathrm{background}$  \\ \hline
      \qsqrbinD          & $236\pm 31$ & $233\pm 32$\\
      \qsqrbinE          & $275\pm 35$ & $363\pm 36$\\
      \qsqrbinF          & $342\pm 39$ & $445\pm 40$\\[0.2pt]
      \hline\hline
    \end{tabular}
  \end{center}
  \vspace*{-0.4cm}
  \caption{The values of fitted signal, $n_\mathrm{signal}$, and background, $n_\mathrm{background}$,
    yields obtained for different bins in \qsq.
    The uncertainties indicated are statistical only~\cite{ATLAS-CONF-2017-023}.} \label{tab:kev}
\end{table*}
\begin{figure}[!tb]
  \begin{center}
    \hspace*{-0.8cm}
    \begin{subfigure}[t]{0.36\textwidth}
      \includegraphics[width=0.99\textwidth]{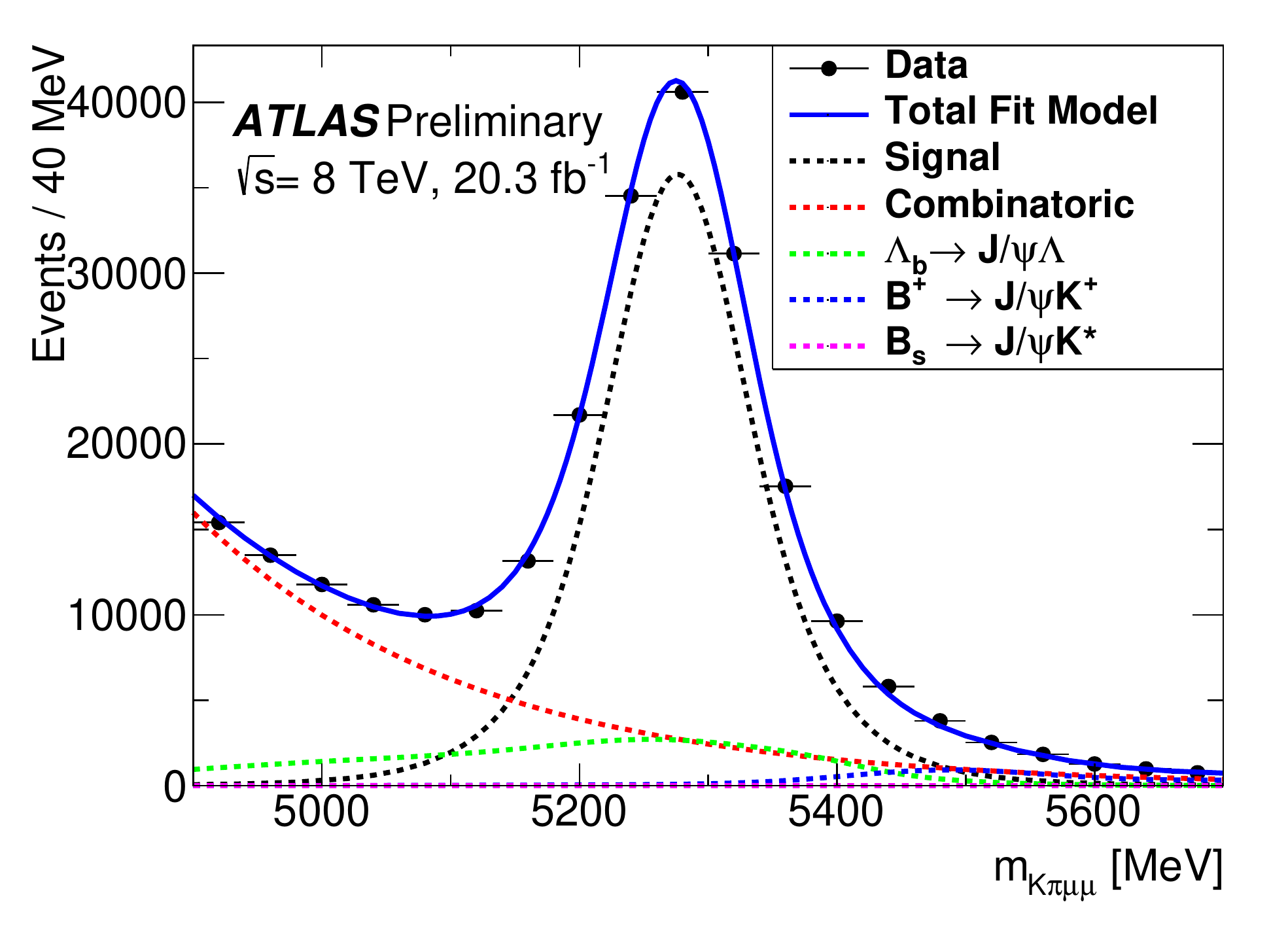}
    \end{subfigure}
    \hspace*{-0.5cm}
    \begin{subfigure}[t]{0.36\textwidth}
      \includegraphics[width=0.99\textwidth]{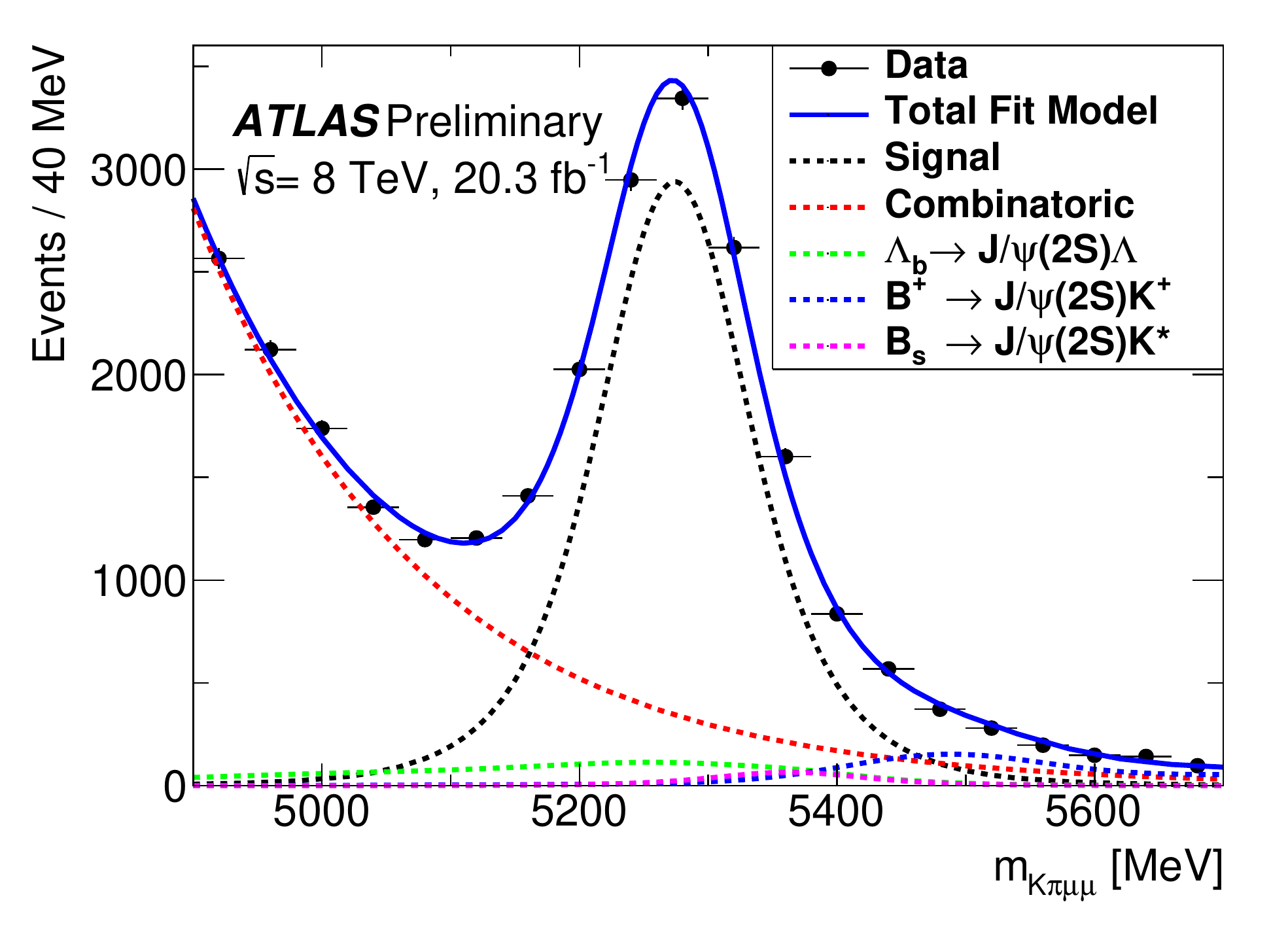}
    \end{subfigure}
    \hspace*{-0.5cm}
    \begin{subfigure}[t]{0.36\textwidth}
      \includegraphics[width=0.99\textwidth]{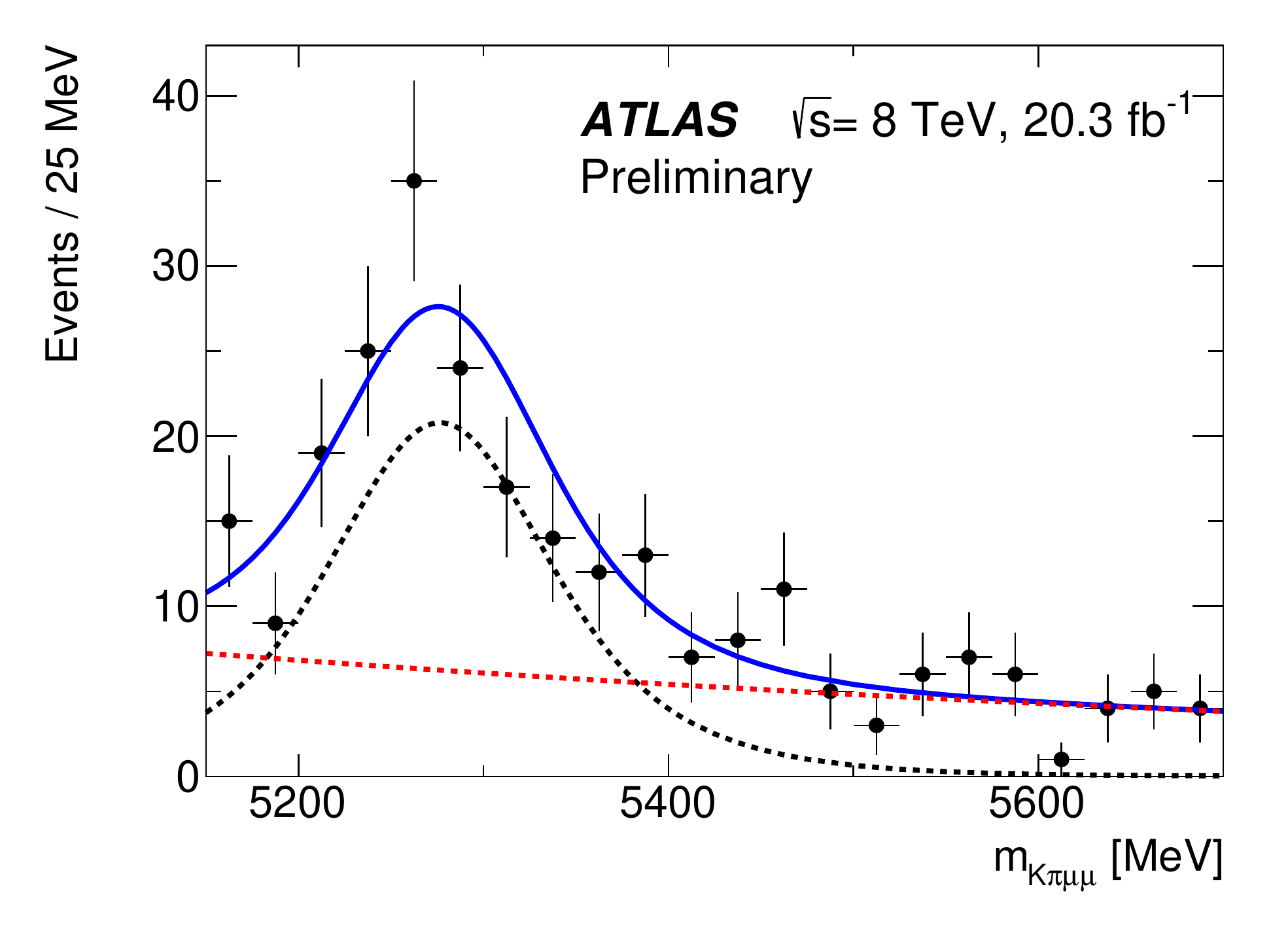}
    \end{subfigure}
    \hspace*{-0.8cm}
    \vspace*{-0.3cm}
    \caption{Control sample fits to the $K\pi\mu\mu$ invariant mass distributions
      for the ({\it{left}}) $K^* \jpsi$ and ({\it{centre}}) $K^*\psitwos$ regions.
      The data are shown as points and the total fit model as the solid lines.
      The dashed lines represent (black) signal, (red) combinatorial background,
      (green) $\Lambda_b$ background, (blue) $B^+$ background and (magenta) $B_s^0$ background.
      {\it{Right:}} distribution of $m_{K\pi\mu\mu}$ obtained for $\qsq\in\qsqrbinA$ \qsqunits.
      The (blue) solid line is a projection of the total p.d.f., the (red) dashed
      line represents the background, and the (black) dashed line represents the
      signal component.  These plots are obtained from a fit using the \Sfive folding scheme~\cite{ATLAS-CONF-2017-023}.
    }
    \label{fig:kctr}
  \end{center}
  \vspace*{-0.5cm}
\end{figure}

\begin{figure}[!bt]
  \begin{center}
    \hspace*{-0.8cm}
    \begin{subfigure}[t]{0.36\textwidth}
      \includegraphics[width=0.99\textwidth]{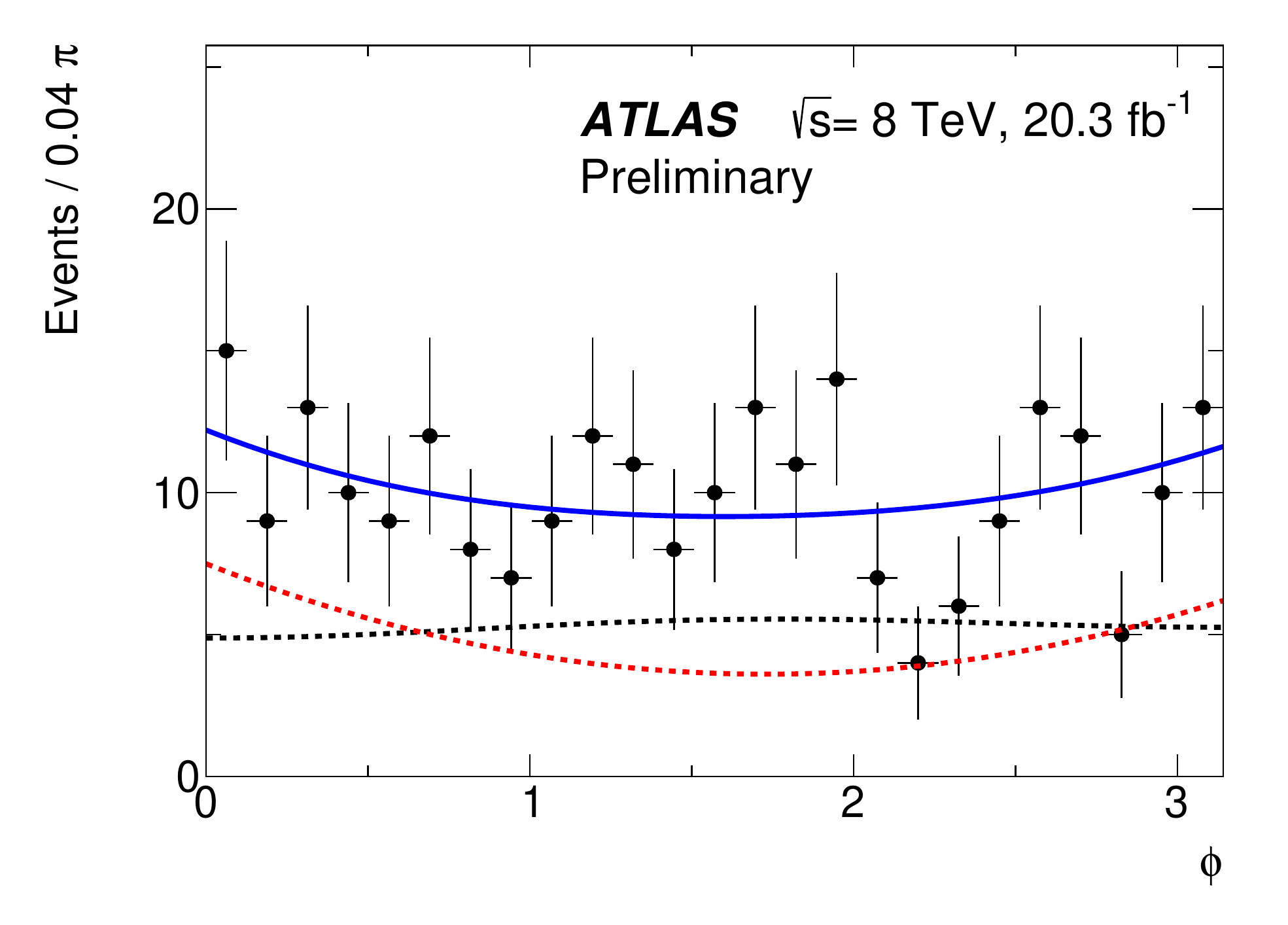}
    \end{subfigure}
    \hspace*{-0.5cm}
    \begin{subfigure}[t]{0.36\textwidth}
      \includegraphics[width=0.99\textwidth]{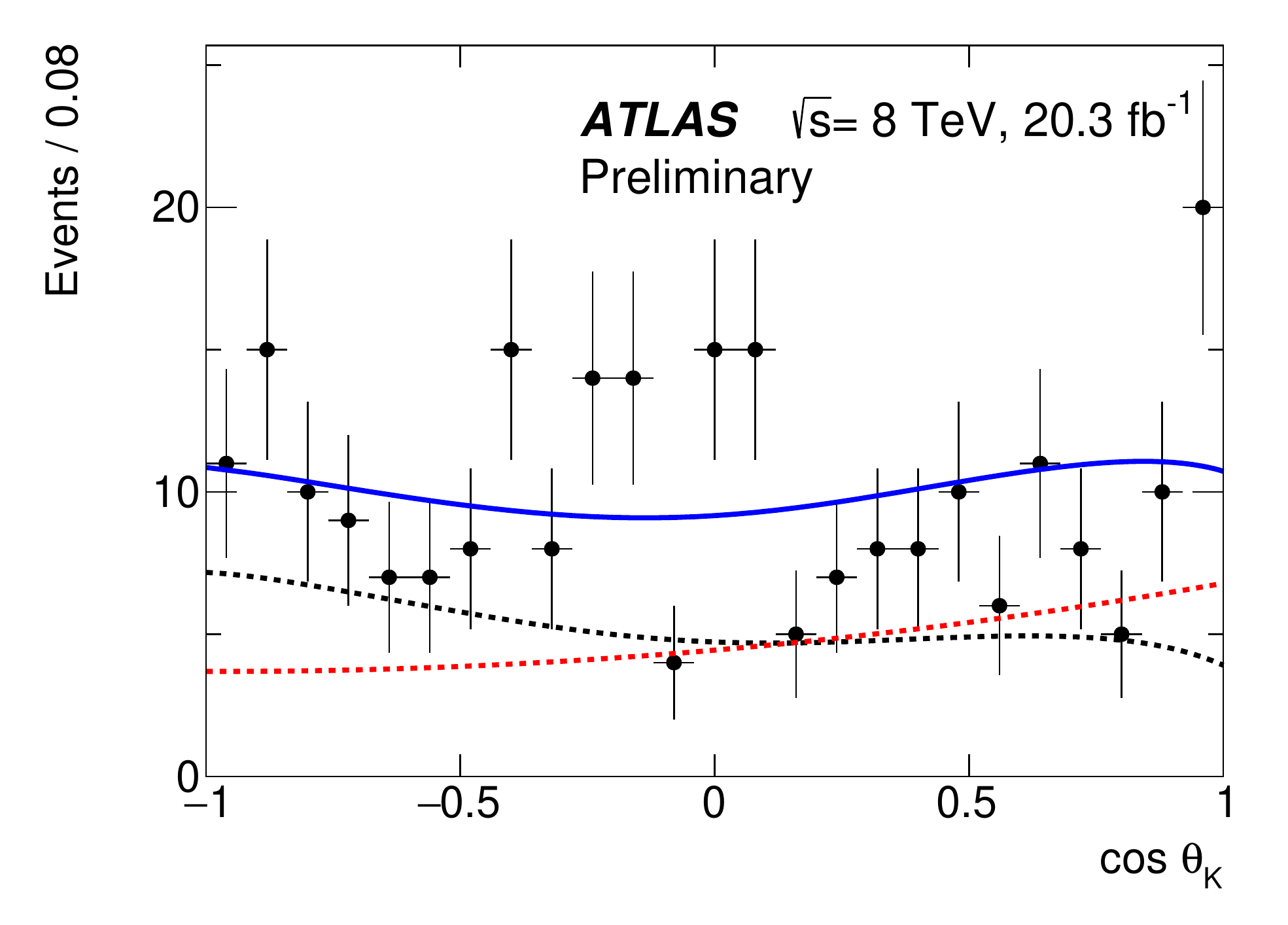}
    \end{subfigure}
    \hspace*{-0.5cm}
    \begin{subfigure}[t]{0.36\textwidth}
      \includegraphics[width=0.99\textwidth]{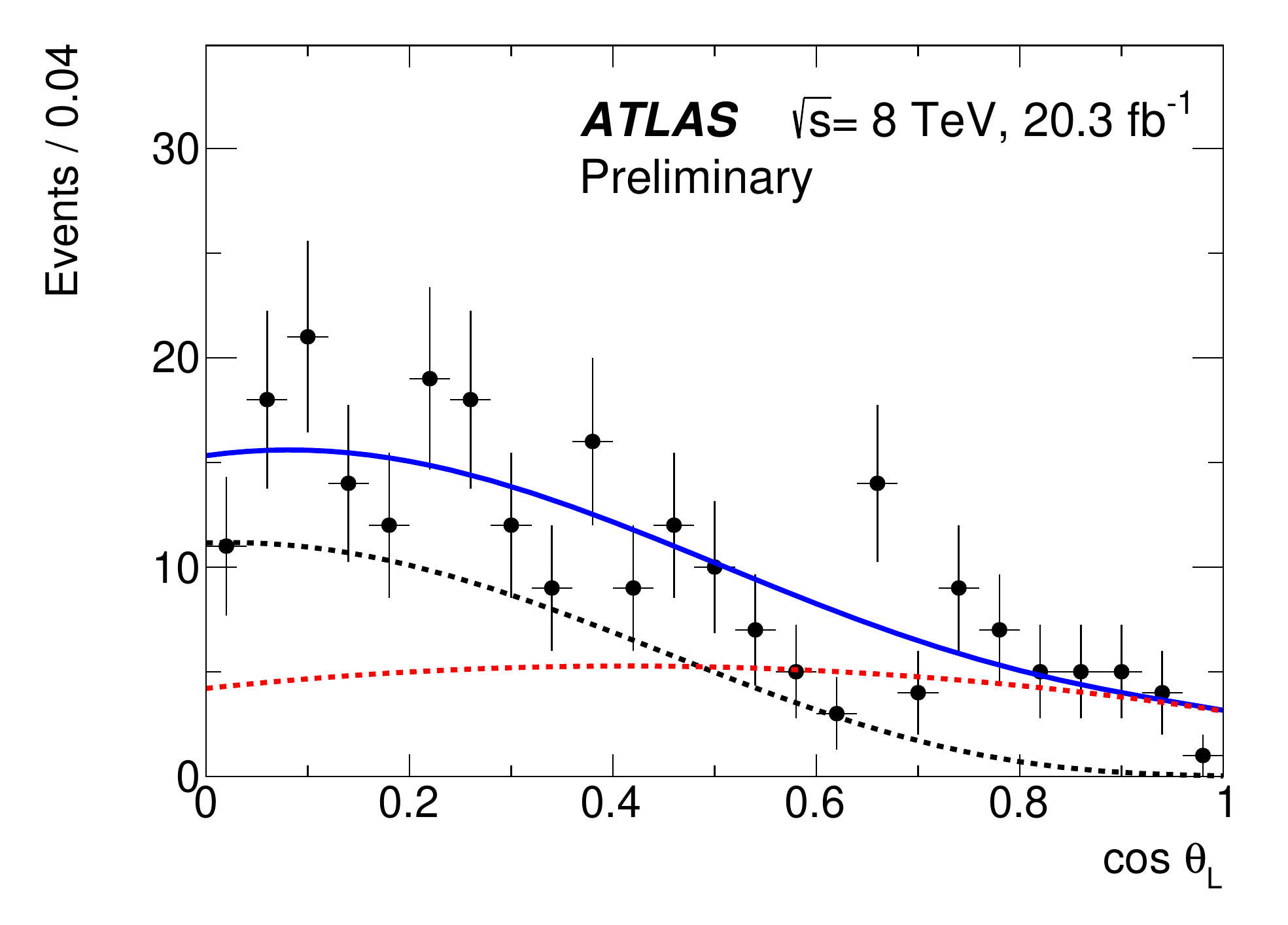}
    \end{subfigure}
    \hspace*{-0.8cm}
    \vspace*{-0.3cm}
    \caption{The distributions of ({\it{left}}) $\phi$, ({\it{centre}}) \ctk, and
      ({\it{right}}) \ctl obtained for $\qsq\in\qsqrbinA$ \qsqunits.
      The (blue) solid line is a projection of the total p.d.f., the (red)
      dashed line represents the background, and the (black) dashed line
      represents the signal component.  These plots are obtained from a
      fit using the \Sfive folding scheme~\cite{ATLAS-CONF-2017-023}.
    }
    \label{fig:ksig}
  \end{center}
  \vspace*{-0.5cm}
\end{figure}

A two-step fit process is performed for the different signal regions in \qsq.
The first step is a fit to the $K\pi\mu^+\mu^-$ invariant mass distribution, using the
event-by-event uncertainty on the reconstructed mass as a conditional variable.
For this fit the signal shape parameters are fixed to the values obtained from
fits to data control samples. The results of the fits are shown in Figures~\ref{fig:kctr} and
the signal yields are given in Tables~\ref{tab:kev}.
A second step adds the (transformed) \ctk, \ctl and $\phi$ variables to the likelihood
to extract \fl and the $S_i$ parameters along with the combinatorial background shapes.
Mass shape parameters and yields are fixed to the results obtained
from the first step. From signal MC samples, the acceptance function is obtained as the
deviation from the generated distribution of \ctk, \ctl, $\phi$ as a result of triggering,
reconstruction and selection of events. The acceptance function multiplies the angular
distribution in the fit.
Figures~\ref{fig:ksig} show for the different \qsq bins the distributions of the variables used
in the fit for the \Sfive folding scheme. Similar sets of distributions are obtained for the
three other folding schemes: \Sfour, \Sseven and \Seight. The results of the angular fits to
the data in terms of the $S_i$ and $P_i^{(\prime)}$ can be found in Table~\ref{tbl:angulartwo}.

\begin{table} [!b]
  \small
  \begin{center}
    \hspace*{-0.8cm}
    \begin{tabular}{l|rrrrrr}\hline\hline
      \qsq [$\qsqunits$]& \multicolumn{1}{c}{\Pone} & \multicolumn{1}{c}{\Pfour} & \multicolumn{1}{c}{\Pfive} & \multicolumn{1}{c}{\Psix} & \multicolumn{1}{c}{\Peight} \\ \hline
      \qsqrbinA        &  $-0.06\pm 0.30\pm 0.10$ & $0.39\pm 0.51 \pm 0.25$ & $0.67\pm 0.26 \pm 0.16$ & $-0.18\pm 0.21\pm 0.04$ & $-0.22\pm 0.38\pm 0.14$ \\
      \qsqrbinB        &  $-0.78\pm 0.51\pm 0.42$ & $-0.96\pm 0.39\pm 0.26$ & $-0.33\pm 0.31\pm 0.13$ & $0.31\pm 0.28 \pm 0.19$ & $0.84\pm 0.32 \pm 0.31$ \\
      \qsqrbinC        &  $0.00\pm 0.47 \pm 0.26$ & $0.81\pm 0.42 \pm 0.24$ & $0.26\pm 0.35 \pm 0.17$ & $0.06\pm 0.27 \pm 0.13$ & $-0.19\pm 0.33\pm 0.07$ \\ \hline
      \qsqrbinD        &  $-0.22\pm 0.26\pm 0.16$ & $-0.38\pm 0.31\pm 0.22$ & $0.32\pm 0.21 \pm 0.10$ & $0.01\pm 0.17 \pm 0.10$ & $0.30\pm 0.26 \pm 0.19$ \\
      \qsqrbinE        &  $-0.17\pm 0.31\pm 0.14$ & $0.07\pm 0.28 \pm 0.18$ & $0.01\pm 0.21 \pm 0.07$ & $0.03\pm 0.17 \pm 0.11$ & $0.18\pm 0.22 \pm 0.16$ \\
      \qsqrbinF        &  $-0.15\pm 0.23\pm 0.10$ & $0.07\pm 0.26 \pm 0.18$ & $0.27\pm 0.19 \pm 0.07$ & $0.03\pm 0.15 \pm 0.10$ & $0.11\pm 0.21 \pm 0.14$ \\
      \hline\hline
    \end{tabular}
  \end{center}
  \vspace*{-0.4cm}
  \caption{The values \Pone, \Pfour, \Pfive, \Psix and \Peight obtained
    for different bins in \qsq.
    The uncertainties indicated are statistical and systematic, respectively~\cite{ATLAS-CONF-2017-023}.
  } \label{tbl:angulartwo}
  \vspace*{-0.5cm}
\end{table}

A total of seven inclusive and eleven exclusive \Bd, \Bs, \Bp and \LambdaB background samples
are studied as systematic uncertainties.
Two additional background contributions are observed in the \ctk and \ctl distributions.
They are also treated as systematic effects given the current statistics.
A peak in \ctk is found at about 1.0 and appears to come from two contributions.
One arises from \Bp\ decays, such as $\Bp\to K^+\mu\mu$ and $\Bp\to \pi^+\mu\mu$,
where an extra track is combined with the hadron to form a fake \kstar. A veto
on events with a three-track invariant mass within a 50~MeV mass window around the nominal \Bp mass
reduces the size of the peak in \ctk.
The second contribution comes from two charged tracks forming a fake \kstar candidate
and it is observed in the $K\pi$ sidebands away from the \kstar meson.
The origin of this source of background is not fully understood.
The background that peaks in \ctl is studied using Monte Carlo simulated events
for the decays $D^0\to K\pi$, $D^+\to K\pi\pi$, $D^+_s\to KK\pi$,
and $D^{*+}_s\to KK\pi$.  Events with an intermediate charm meson,
$D^0$, $D^\pm_{(s)}$ and $D^{*+}_s$ are found to accumulate around 0.7 in $|\ctl|$.
A 30~MeV wide veto window about the reconstructed charm meson mass can be applied
to eliminate this background.

The main systematic uncertainties come from backgrounds, mainly the fake \kstar background
and the background arising from partially reconstructed $B \to D^0/D^+/D^+_s/D^*_s X $ decays.
Other systematic uncertainties in decreasing order of importance are coming from the background
p.d.f.\ shape, the acceptance function, the combinatorial background, tracking alignment,
knowledge of the magnetic field, bias from the maximum likelihood estimator, $p_{\text{T}}$ spectrum
of $B$ candidates, and scalar contributions from non-resonant $K\pi$ transitions,
The total systematic uncertainties for the $P_i^{(\prime)}$ parameters
fitted are presented in Table~\ref{tbl:angulartwo}.

\begin{figure}[!ht]
  \begin{center}
    \resizebox{\columnwidth}{!}{
      \hspace*{-0.9cm}
      \includegraphics[width=0.53\columnwidth]{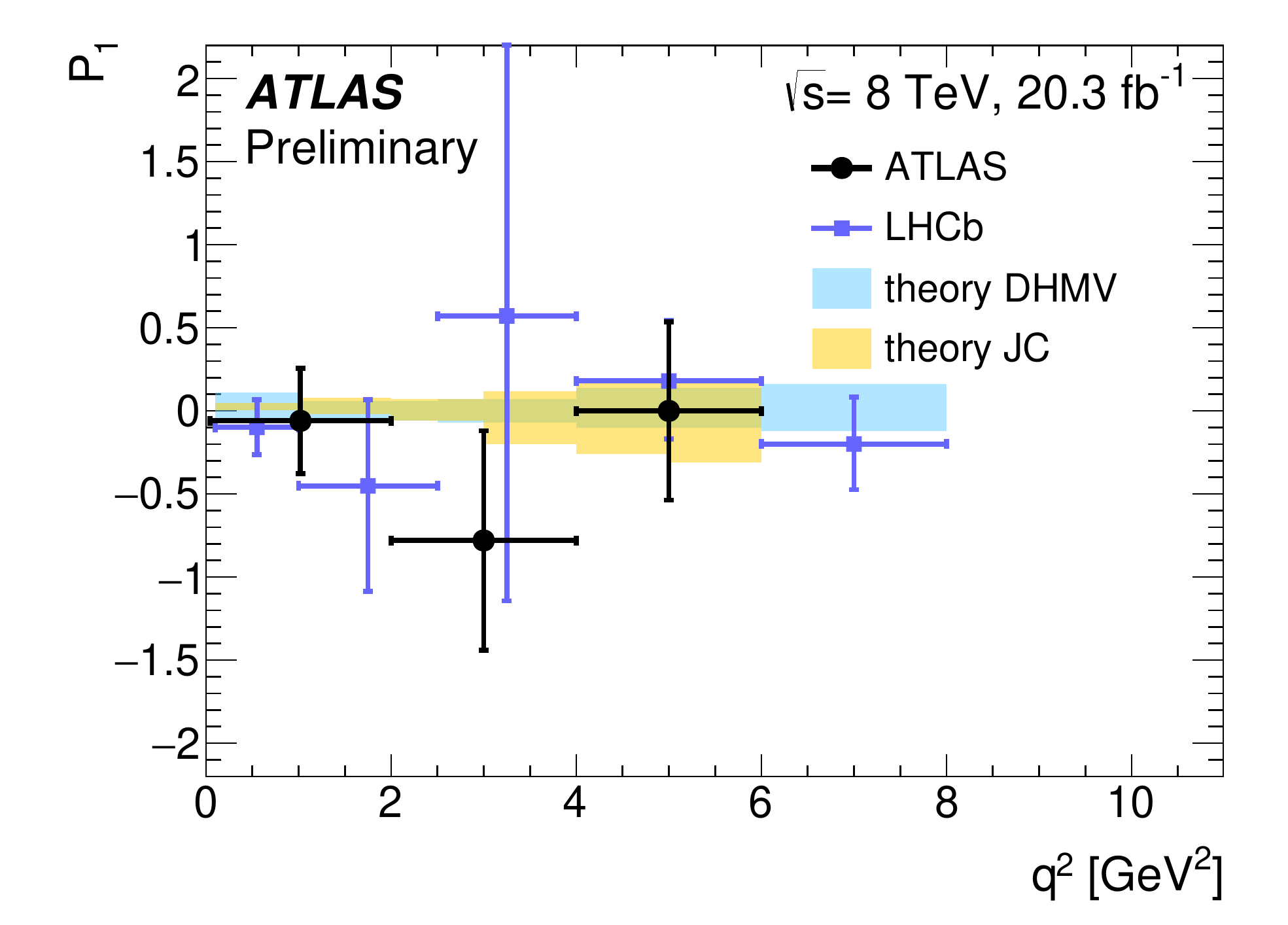}
      \hspace*{-0.6cm}
      \includegraphics[width=0.53\columnwidth]{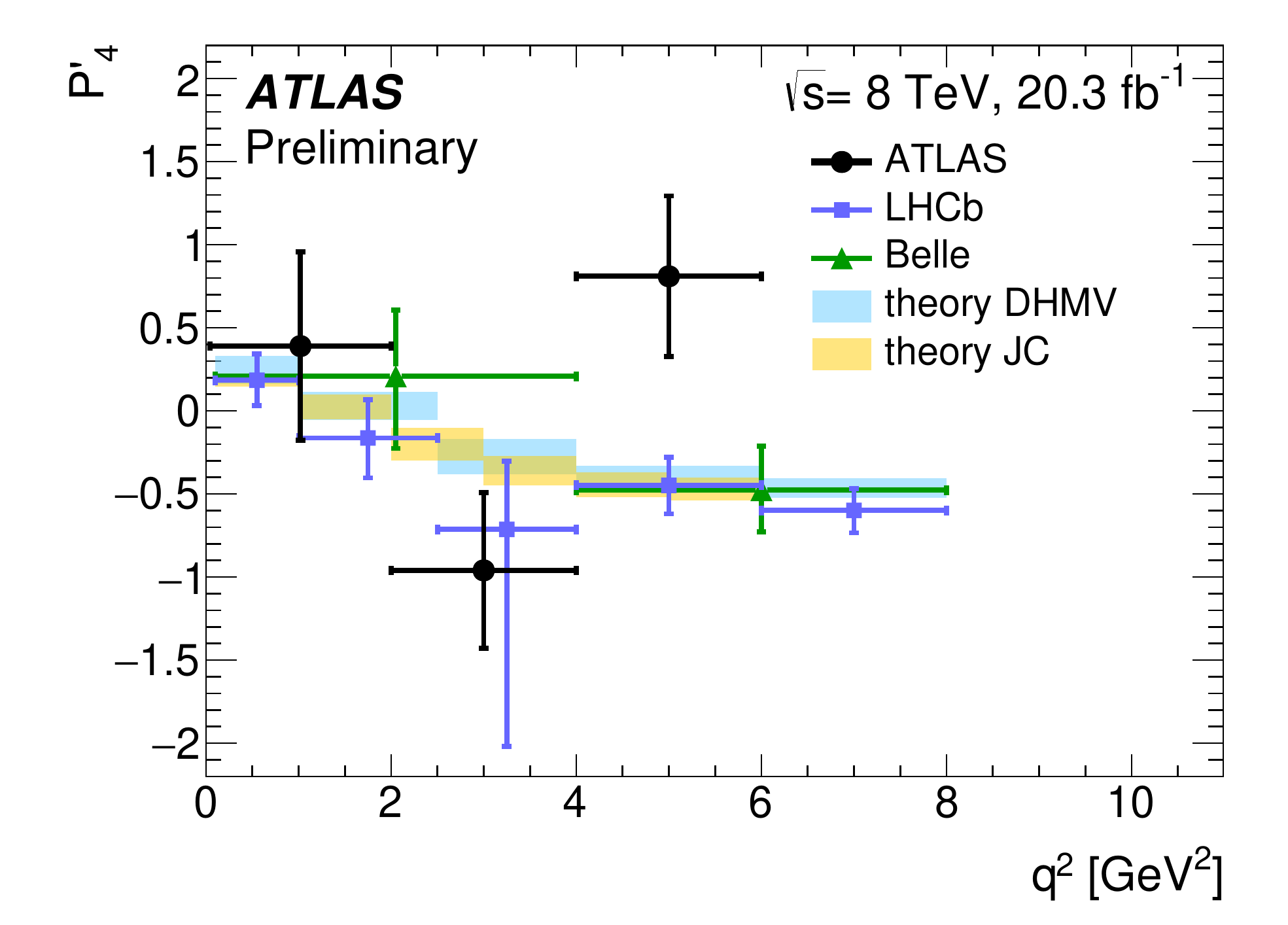}
      \hspace*{-0.6cm}
      \includegraphics[width=0.53\columnwidth]{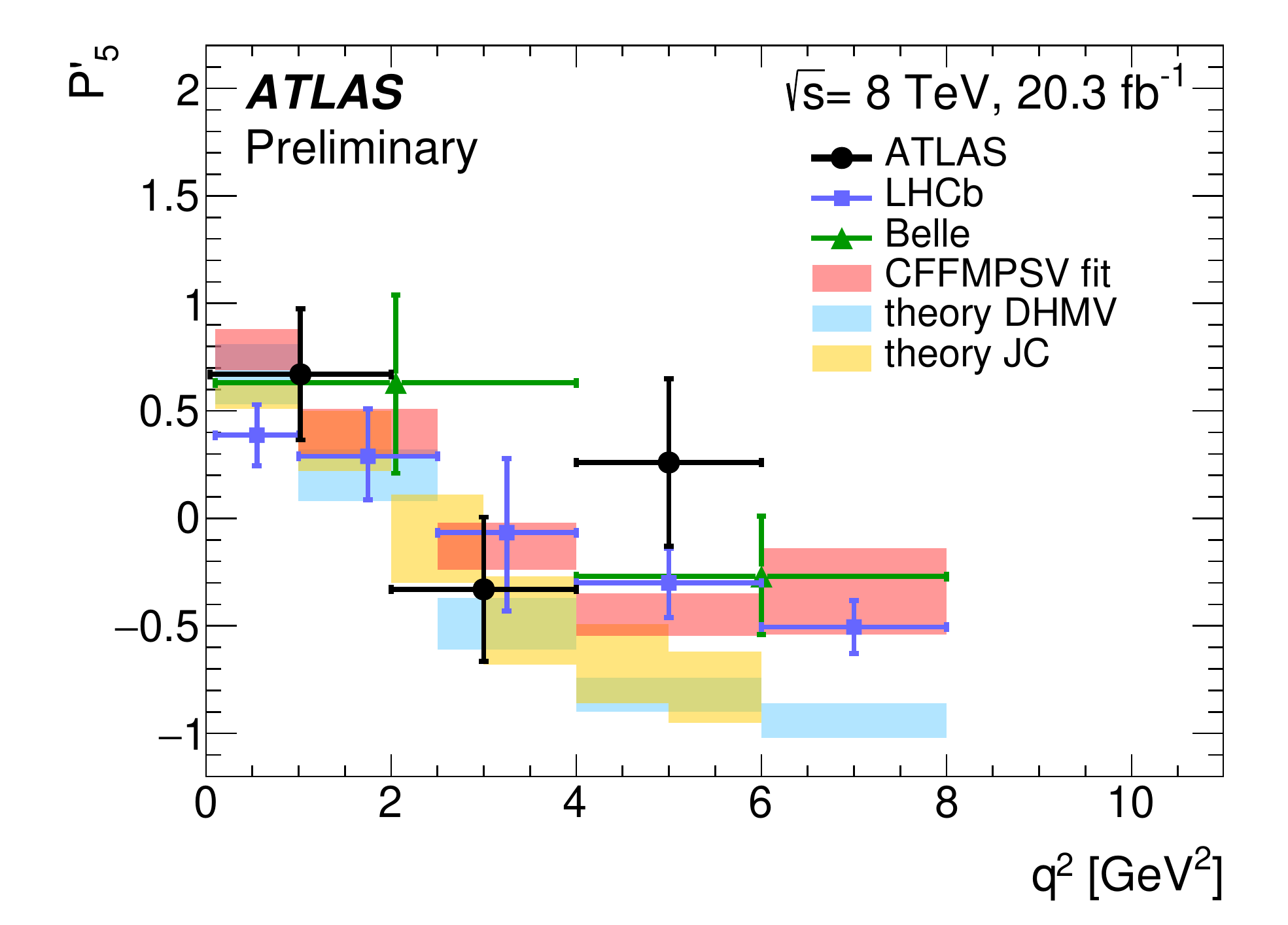}
      \hspace*{-0.8cm}
    }
    \resizebox{0.72\columnwidth}{!}{
      \includegraphics[width=0.56\columnwidth]{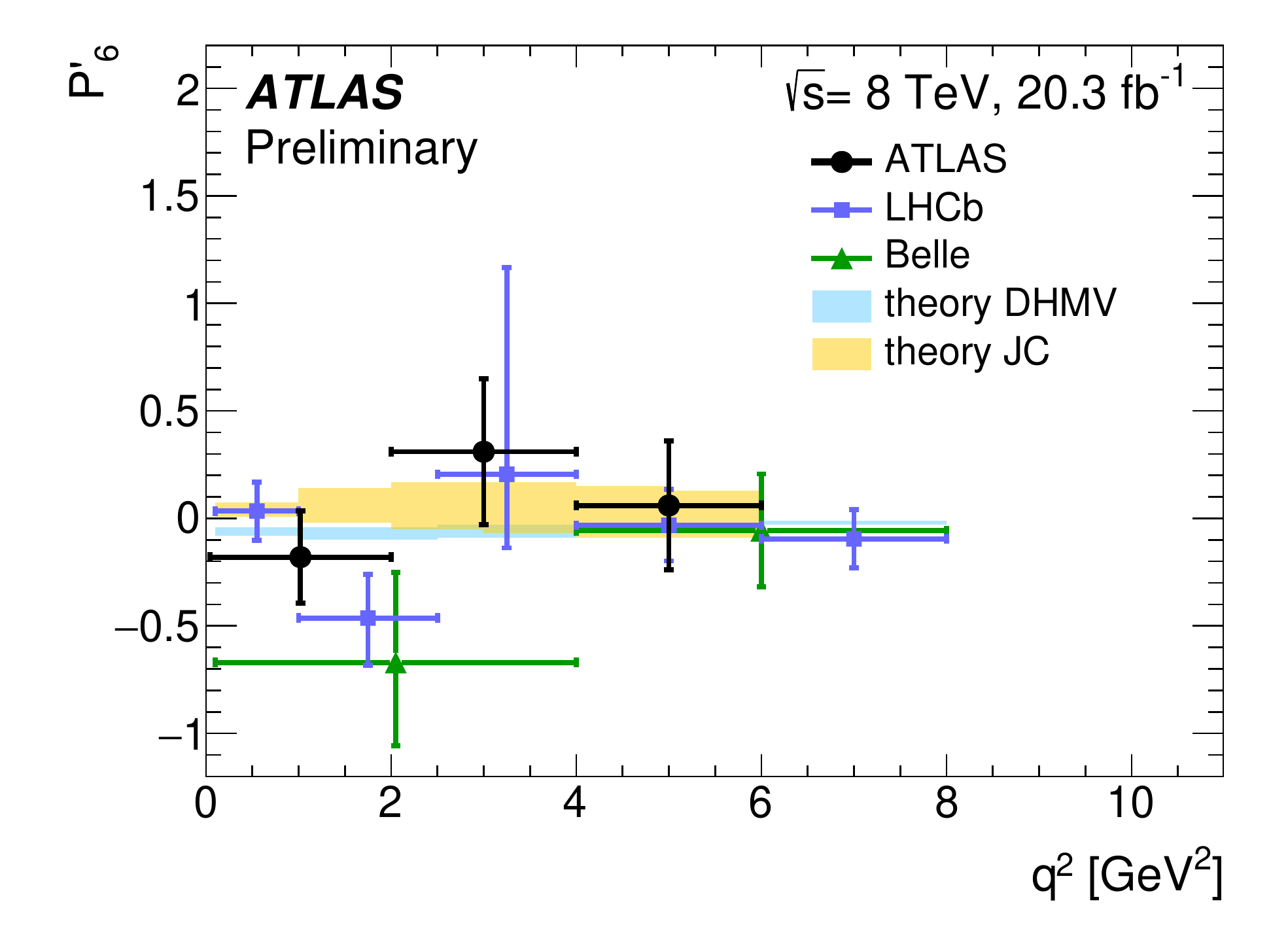}
      \hspace*{-0.6cm}
      \includegraphics[width=0.56\columnwidth]{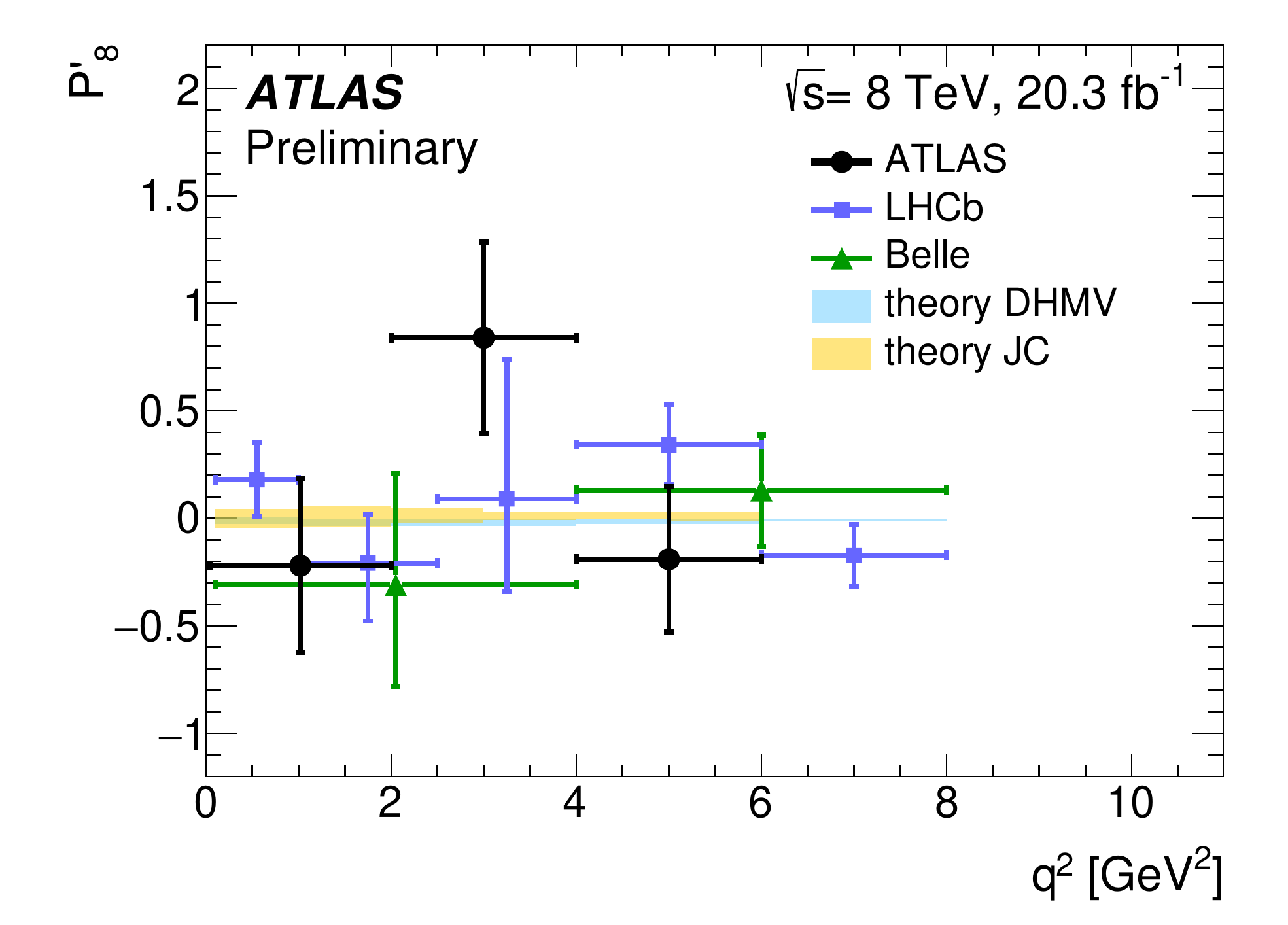}
    }
  \end{center}
  \vspace*{-0.5cm}
  \caption{The measured values of \Pone, \Pfour, \Pfive, \Psix, \Peight compared with predictions from
    the theoretical calculations discussed in the text.
    Statistical and total uncertainties are shown for the data, i.e. the inner mark
    indicates the statistical uncertainty and the total error bar the total uncertainty~\cite{ATLAS-CONF-2017-023}.
    \label{fig:kres}}
\end{figure}

The results of theoretical approaches of Ciuchini et al. (CFFMPSV)~\cite{Ciuchini:2015qxb},
Descotes-Genon et al. (DHMV)~\cite{Descotes-Genon:2014uoa}, and J\"ager and Camalich (JC)~\cite{Jager:2012uw,Jager:2014rwa}
are shown in Figure~\ref{fig:kres} for the $P^{(\prime)}$ parameters.
With the exception of the \Pfour and \Pfive measurements in $\qsq\in\qsqrbinC$ \qsqunits
and \Peight in $\qsq\in\qsqrbinB$ \qsqunits there is good agreement between theory and measurement.
The deviation observed for \Pfour (\Pfive) is consistent with the one reported by
the LHCb Collaboration~\cite{Aaij:2015oid}, and it is approximately 2.5 (2.7) standard
deviations away from the calculation of DHMV. The deviations are less significant for the other
theoretical approaches.
All measurements are found to be within three standard deviations of the range covered
by the different predictions. Hence, including experimental and theoretical uncertainties,
the measurements presented here are found to be in accordance with the expectations
of the SM contributions to this decay.

\bibliographystyle{JHEP}
\bibliography{bona_lhcp17_atlas}{}

\end{document}